\documentclass[acmsmall]{acmart}
\usepackage{soul}
\usepackage{tabularx}
\usepackage{enumitem}
\usepackage{listings}

\setcopyright{cc}
\setcctype{by-sa}
\acmDOI{10.1145/3748612}
\acmYear{2025}
\acmJournal{PACMHCI}
\acmVolume{9}
\acmNumber{6}
\acmArticle{GAMES017}
\acmMonth{10}
\received{2025-02-19}
\received[accepted]{2025-07-24}

\begin{document}

\title[Democratizing Game Modding with GenAI: A Case Study of StarCharM]{Democratizing Game Modding with GenAI: A Case Study of StarCharM, a Stardew Valley Character Maker}

\author{Hamid Zand Miralvand}
\orcid{0009-0007-6223-0791}
\affiliation{%
  \institution{Polytechnique Montréal}
  \city{Montréal}
  \country{Canada}
}
\email{hamid.zand-miralvand@polymtl.ca}

\author{Mohammad Ronagh Nikghalb}
\orcid{0009-0007-9889-9024}
\affiliation{%
  \institution{Polytechnique Montréal}
  \city{Montréal}
  \country{Canada}
}
\email{mohammad.ronagh-nikghalb@polymtl.ca}

\author{Mohammad Darandeh}
\orcid{0009-0000-9470-2924}
\affiliation{%
  \institution{Polytechnique Montréal}
  \city{Montréal}
  \country{Canada}
}
\email{mohammad.darandeh@polymtl.ca}

\author{Abidullah Khan}
\orcid{0000-0002-2131-9497}
\affiliation{%
  \institution{Polytechnique Montréal}
  \city{Montréal}
  \country{Canada}
}
\email{abid-ullah.khan@polymtl.ca}

\author{Ian Arawjo}
\orcid{0000-0001-8910-0822}
\affiliation{%
  \institution{Université de Montréal}
  \city{Montréal}
  \country{Canada}
}
\email{ian.arawjo@umontreal.ca}

\author{Jinghui Cheng}
\orcid{0000-0002-8474-5290}
\affiliation{%
  \institution{Polytechnique Montréal}
  \city{Montréal}
  \country{Canada}
}
\email{jinghui.cheng@polymtl.ca}

\renewcommand{\shortauthors}{Zand Miralvand, Nikghalb, Darandeh, Khan, Arawjo, and Cheng}

\begin{abstract}
Game modding offers unique and personalized gaming experiences, but the technical complexity of creating mods often limits participation to skilled users. We envision a future where every player can create personalized mods for their games. To explore this space, we designed StarCharM, a GenAI-based non-player character (NPC) creator for Stardew Valley. Our tool enables players to iteratively create new NPC mods, requiring minimal user input while allowing for fine-grained adjustments through user control. We conducted a user study with ten Stardew Valley players who had varied mod usage experiences to understand the impacts of StarCharM and provide insights into how GenAI tools may reshape modding, particularly in NPC creation. Participants expressed excitement in bringing their character ideas to life, although they noted challenges in generating rich content to fulfill complex visions. While they believed GenAI tools like StarCharM can foster a more diverse modding community, some voiced concerns about diminished originality and community engagement that may come with such technology. Our findings provided implications and guidelines for the future of GenAI-powered modding tools and co-creative modding practices.
\end{abstract}



\maketitle

\section{Introduction}
Video game modification, or modding in short, is a hands-on practice carried out by game players or fans (instead of the game developers themselves) to change and enhance the game content~\cite{poor2014computer, scacchi2010computer}. Nowadays, this practice is popular among gamers and supported by large communities. For instance, Nexus Mods\footnote{\url{https://www.nexusmods.com}}, one of the largest game mod hosting websites, hosts more than 600,000 mods for more than 3,000 games, created by hundreds of thousands of authors and enjoyed by millions of players. There are diverse forms of game modding~\cite{scacchi2010computer}, including user interface customization, game conversions (i.e., adding or changing in-game characters, objects, mechanics, etc.), and machinima and art mods. Research shows that modders are driven primarily by a love for the game and desires to enhance their gameplay experience~\cite{poor2014computer, unger2012modding}, express themselves artistically and creatively, and feel a sense of belonging to a community~\cite{tancred2023understanding, SotamaaModdersMotivation}. In sum, modding gives players and mod creators more ways to experience the game and express themselves.

In spite of its popularity and benefits, mod creation, however, is not a practice that is easily attainable for everyone. Creating a mod often requires extensive, game-specific technical skills, such as familiarity with a particular scripting language, graphic design tool, or software development kit, as well as an understanding of community norms, which can pose significant barriers for newcomers~\cite{OfModsandModders,kow2009culture}. Some modding practices also require reverse engineering skills to unravel the game's design and inner working details, often through hacking-like strategies~\cite{scacchi2011modding}. Although some game engines, such as iD Tech and Unreal, actively support modding by providing documentation and infrastructure, learning the necessary workflows and programming languages can still be daunting for novices~\cite{ModdingEngine}. As such, there appears to be a layered hierarchy in modding communities, where at the center are dedicated mod creators, surrounded by players who actively use mods, who are surrounded further by the vast majority of regular players who may have never encountered modding. While mod users and non-modders are not necessarily casual players, they often have limited power to customize their gaming experience and end up becoming the ``consumers'' of the mod content and the game, relying upon the expertise and creativity of others. As a result, they miss the opportunity to enjoy their favorite games in unique and personalized ways and to showcase their creativity. 

The advent of generative artificial intelligence (GenAI) promises to improve the situation, empowering people with less technical skills to implement software through natural language expressions of their intent~\cite{panchanadikar2024m,nguyen2024beginning,sweetser2024large}. Indeed, previous studies have explored the use of GenAI in game design and development, leveraging its capabilities to generate game content such as narratives~\cite{SteenstraHealthEducationLLM}, levels~\cite{irfan2019evolving}, quests~\cite{AshbyQuestGenerator, RPGQuestGenerator}, visual assets~\cite{karp2021automatic}, and game rules~\cite{TorriLotteryAndSprint}. However, while GenAI has significant potential to enhance the modding process, there is no empirical research specifically studying this question. One might presume that GenAI can ease the burden of content creation and programming efforts for game modding; however, the usage of GenAI might also raise thorny questions of authenticity, authorship, and ethics~\cite{bozkurt2024genai, sun2024generative}. In this work, we address this area and aim to explore how players perceive GenAI-empowered tools in transforming the game modding landscape. While modding encompasses a wide range of practices, we focus specifically on non-player character (NPC) creation in the game \textit{Stardew Valley} and investigated the design and user perception of a tool called \textbf{StarCharM} (\underline{Star}dew Valley \underline{Cha}racter \underline{M}aker) for adding NPCs in the game.

Stardew Valley is a farming simulation role-playing game where players manage an inherited farm. By partaking in a wide range of activities, such as mining, fishing, foraging, and farming, players work to restore the farm and enjoy a country life. As they go about their daily tasks, they can interact with various villagers (NPCs), build relationships, and even marry and start a family. The game places a strong narrative emphasis on NPC interactions, emotional arcs, and social routines. First released in 2016, Stardew Valley is widely considered to be one of the most successful independent games of all time, with continual updates and a large fanbase\footnote{\url{https://www.stardewvalley.net/press/}}. 

We chose Stardew Valley both for this popularity and because it has an active modding community~\cite{stardewvalleywikiModdingCommunityStardew} with strong modding support~\cite{stardewvalleywikiModdingIndexStardew}. Its modding community is exceptionally active, diverse, and welcoming to novice creators (including the authors of this paper), with accessible documentation and widespread use of tools like SMAPI and Content Patcher. Stardew Valley supports a wide range of modding practices, including retexturing visual assets, adding automation tools (e.g., for more efficient farming), redesigning in-game objects to streamline gameplay, and adding new NPCs and narrative arcs. Among these mod types, we focused on NPC modding in this game because of two reasons. First, NPC modding in Stardew Valley provides a rich canvas for personal storytelling, offering the flexibility to cater to different player preferences (as mentioned above and see more in Section~\ref{sec:exploration}). The NPCs in the game essentially provide a rich setting and lore to the players, offering unique backstories, personalities, and daily routines. As such, a diverse array of players can use NPC modding as a means of personal expression and to enrich their personal experiences in the game~\cite{cook2024mod}. Second, while technically challenging, NPC mods in Stardew Valley have a structured design (through configuration files that define the NPC's traits, schedules, dialogue blocks, and gift preferences) that makes them a well-scoped and modular starting point for exploring the complex issue of GenAI-supported mod creation. This combination of narrative richness and modularity motivated our focus. Through this focused lens, we seek to derive insights into how GenAI-based NPC-creation modding tools influence the experiences, expectations, and values of diverse players and at the same time, aim to surface broader implications for the future of GenAI-assisted modding tools and ecosystems.

The design of StarCharM is informed by both our own experience playing and modding Stardew Valley and our preliminary investigation of the existing NPC mods of the game and the surrounding modding community. Unlike traditional modding, which requires coding and configuration knowledge, StarCharM enables NPC creation through an iterative and progressive process of describing, selecting, and tweaking character traits. This process aims to make modding accessible to players who lack programming skills or familiarity with the mod structure.

We conducted a user study of StarCharM with ten Stardew Valley players who had varied modding backgrounds to understand how they would perceive the potential impact of GenAI on the NPC modding process, their gaming experience, and the player and modding communities they are members of. Results indicated that StarCharM offered a structured process for NPC mod creation, enabling participants to craft unique characters and exercise creative control. However, the tool's AI integration raised concerns among participants about potentially diminishing the value of originality and craftsmanship in modding and weakening community bonds. Additionally, while participants felt StarCharM served as an entry point for new modders, it fell short of supporting rich, complex content. Overall, this work contributes to the understanding of the potential and complexity of using GenAI-infused tools to support the modding process. We establish insights about how these types of tools can shape player and modder expectations and impact the modding ecosystem. The StarCharM tool, our study results, as well as our reflection on our design and study process, also provide implications for designing future GenAI-powered modding tools to allow a wide range of users to engage in this practice.

\section{Related Work}
Our study is related to previous work in three main areas: (1) the practice of game modding, (2) the application of AI for content generation in games, and (3) game character design with the support of AI. We briefly review each body of literature below.

\subsection{Game Modding}
Game modding refers to the practice of making alterations and additions to preexisting games by the game players~\cite{SotamaaModdersMotivation}. This practice has grown to be an important aspect of both video game culture and the gaming industry~\cite{OfModsandModders}. Previous work has shown that modding not only fosters creativity but also serves as valuable practice for aspiring developers and designers to learn about game development~\cite{LearningWithModding}. Previous research has also identified that the modding practice had the potential to enable the representational experience of players who had diverse identities~\cite{cook2024mod}. Players today can modify various aspects of a video game, including environments, characters, quests, and even game rules~\cite{MaPlayersCollaboratation, scacchi2010computer}. Mod creators are driven by a variety of factors when developing mods~\cite{SotamaaModdersMotivation, poor2014computer, tancred2023understanding, OfModsandModders}. For example, \citet{OfModsandModders} identified three main factors motivating mod creators: (1) to express their artistic and creative passions, (2) to enhance their enjoyment of the game, and (3) a way to obtain jobs in game companies. In addition to individualistic motivations, mod creators and users also engage in modding for the social aspects, which include community engagement, socialization, and collaboration~\cite{SotamaaModdersMotivation, tancred2023understanding}. The modding community is considered to have an open-source attitude that everyone can edit and reuse the mods~\cite{unger2012modding}. Modders share a strong sense of community and enjoy helping each other in a collaborative environment~\cite{poor2014computer}. In response to the modding needs, game developers are increasingly offering tools and editors to facilitate user-generated content creation~\cite{ModdingEngine,Modulith}. For example, \citet{ModdingEngine} found that many game engines provided unique features to support game modding, which had varied implications for players (e.g., types of mods that can be used) and mod creators (e.g., development effort). However, these tools often require a basic understanding of game engines or programming skills. Building on top of this literature, our study explores how players and game modders perceive AI-enabled tools that can help overcome this barrier and encourage a wider range of players to create game mods. 

\subsection{AI Game Content Generation}
Traditionally described as procedural content generation (PCG)~\cite{summerville2018procedural} and largely influenced by the advancement of AI technologies like large language models (LLMs)~\cite{hong2024game}, AI content generation has increasingly shaped the video game industry. Previous work on PCG methods focused on automatically creating game content, such as game levels and characters~\cite{shaker2016procedural, summerville2018procedural}. Although new methods of PCG using deep learning have greatly improved the quality of the generated content~\cite{liu2021deep}, there were still issues regarding the diversity and usability of the generated content~\cite{summerville2018procedural}.

LLMs provided unparalleled flexibility in content generation, allowing for the development of dynamic video game content that adapts and evolves based on user input~\cite{buongiorno2024pangea}. These AI-driven advancements allow developers to create more engaging and varied experiences tailored to individual player preferences. Several recent studies have explored the use of large language models (LLMs) in content generation, intending to adapt and improve this technology for professionals in the game industry. For example, \citet{SteenstraHealthEducationLLM} investigated the use of LLM-generated narrative games for health education. \citet{TorriLotteryAndSprint} also developed a board game creation methodology called Lottery and Sprint, which employs AutoGPT and the Design Sprint framework to generate board games. In addition to generating the entire game, LLMs have been utilized to generate specific game contents, such as levels with adjustable difficulty in games~\cite{ToddLLMLevelGenerator,sudhakaran2024mariogpt}, quests for role-playing games~\cite{RPGQuestGenerator,AshbyQuestGenerator}, and game narratives~\cite{buongiorno2024pangea}. For example, \citet{AshbyQuestGenerator} created an LLM-based framework that leveraged a game-specific knowledge base to dynamically generate quests and NPC dialogues based on a line of dialogue input by the player. \citet{jennings2024s} also proposed GROMIT, which is an LLM-based runtime behavior generation system that can be triggered by player action. The system can integrate knowledge of the environment into the LLM prompt, generate runtime code based on player input, and integrate it into a large-scale game system.

With the growing usage of LLMs across various fields, recent surveys have systematically reviewed the application of LLMs in game development and design. Gallotta et al.~\cite{gallotta2024large} provided a comprehensive road map of how LLMs are being used for tasks ranging from narrative generation to game testing and player modeling. Yang et al.~\cite{yang2024gpt} conducted a scoping review of GPT-based applications from 2020 to 2023, highlighting trends in content generation, dialog systems, and non-player character (NPC) interaction. While these works reveal a rapidly expanding field, our study contributes an underexplored perspective: using LLMs to support character modding by non-technical players through accessible co-creation tools. Rather than focusing on AI as a generator or agent, we frame StarCharM as a scaffold for human creativity in the modding context. As such, we do not aim to directly investigate character generation technologies. Instead, we used a widely available technology to explore how such a type of technology may impact the game mod creation experience and the modding communities.

\subsection{AI-Supported Game Character Design}
Regardless of the role AI plays, game character design has long been investigated as a vital aspect of games, studied from psychological, narrative, and interaction design perspectives. Prominent work done by \citet{isbister2022} offered foundational principles for designing emotionally resonant game characters, emphasizing personality traits, social cues, and player empathy. \citet{lankoski2008} framed characters as central agents of gameplay conflict and progression, while \citet{vandewalle2024} proposed a framework for understanding how characterization emerges through narrative, mechanics, and player interpretation. Other studies have examined how players can co-create character identity and responsiveness through interaction~\cite{kleinsmith2013} and how character diversity plays a role in player inclusion and representation~\cite{to2018diversity}.

More closely related to our study, recent efforts have explored character design using AI. A recent literature review has found that generative AI is transforming the entire game character design workflow, supporting character conceptualization, visual design, props and garments creation, as well as behavior and animation design~\cite{wu2025systematic}. Various tools and techniques were proposed. For instance, \citet{QinCharacterMeet} created CharacterMeet, a tool for creative writers, including game designers, to ideate characters through conversations with an LLM-based chatbot. After the character's basic attributes, the backstory, and a conversation context are provided, users can chat with the character through voice or text to ideate the character and plot details. The users can also customize the character's voice and appearance to visualize the character in a scene. The user study results indicated that the tool facilitated the participants' iterative character-creation processes. Directly targeting character design in games, \citet{ling2024sketchar} proposed Sketchar to assess the interaction between game designers and GenAI-based character design support tools. Using Sketchar, character designers can input the conceptual character specifications and iterate on the generated character profile and appearance. Their user study results indicated that Sketchar facilitated the refinement of design details and could be integrated into real-world character design workflows; character designers without artistic backgrounds also found the Sketchar process to be more expressive and valuable. 
Focusing on creating assets for developers, \citet{SunText2AC} also developed Text2AC, a framework for generating 2D character images from natural language descriptions.

In our study, we drew inspiration from the previous work for the design of StarCharM. However, our main focus is to use this character design tool as a probe to investigate how players perceive the impact of such a tool on game modding and the modding communities. We build on this literature by designing a tool that scaffolds character creation through LLMs, allowing players to specify and iterate on traits such as personality, dialogue, routines, and values. While previous systems focus on authoring characters manually or with system support, StarCharM contributes a co-creative workflow that blends natural language input and iterative refinement, enabling a wider range of players to engage in rich character creation processes grounded in both narrative and gameplay design.

\section{Exploration of the Existing NPC Mods of Stardew Valley}
\label{sec:exploration}
To understand the general goals and purposes for modders to create new NPCs in Stardew Valley and to inform the design of StarCharM, we analyzed the existing Stardew Valley NPC mods on Nexus Mods (nexusmods.com), the most popular website for Stardew Valley mods hosting and sharing. We adopted the following steps to select mods for analysis; data collection happened in June 2024. First, we filtered the Stardew Valley mods with the category of ``New Characters'' on Nexus. We ordered the returned 542 new NPC mods by the number of user ``Endorsements'', which indicates users' appreciation for a mod; this allowed us to focus on popular mods that have attracted a wide range of users. Then, from the ordered list, we examined each mod to include ones that were made in English and removed expansion mods because they included many elements alongside new NPCs. We then selected the top 30 mods on our list for analysis. The analysis process for each mod included reading the developer's description, reviewing the mod's JSON files to determine interactions with other existing characters, and reading player comments on the mod's webpage. We extracted information about each mod in various aspects, such as the number of new NPCs they contained, whether those NPCs had personalities and backstories, and whether they were connected to the existing Stardew Valley NPCs, as well as the primary goals set by their developers for creating the mods. Based on the extracted information, we categorized the mods into the following three main groups.

\textbf{Casual Mods ($N=21$)}: These mods feature characters and content designed to add enjoyable, lighthearted experiences to the game. Characters in these mods often have simple backstories and engage players with simple quests that involve helping with their work or running errands. Some of these characters have simple connections to the existing characters to further improve their personality. For example, in the Mister Ginger mod, the newly added cat NPC would simply follow Jas, one of the existing characters, to add a fun flavor to the game. Some mod developers even bring characters from other games or animated series into Stardew Valley, which can be seen as a way to merge different fan bases, provide nostalgic value, and/or offer players the novelty of interacting with beloved characters in a new environment. For instance, the Satoru Gojo mod adds a character from the popular anime Jujutsu Kaisen with the same personality that he had in the anime series.
    
\textbf{Impactful Narrative Mods ($N=3$)}: The second category consists of more serious mods, which typically introduce characters with deep, often tragic backstories addressing issues such as alcoholism and depression. These mods aim to make an emotional impact on players by narrating stories of characters who have overcome hardships and problems. The characters in this type of mod have a moderate amount of interactions with existing characters to give them more depth. By presenting these heavier narratives, developers of these mods seek to create a more thought-provoking and immersive experience. A good example of this category would be the ``Always Raining in the Valley'' mod, which portrays three characters with immersive and tragic backstories. The characters each tell the player their stories and mistakes in the past, and how they changed their lives and got over their problems.
    
\textbf{Story Augmentation Mods ($N=6$)}: The final category includes mods that add characters referenced by the NPCs within the original game to enrich the existing story and lore. They were created in response to the intentionally open-ended design of the existing NPCs in the game, which invites players to imagine what could happen beyond the official dialogue and events. These added characters often have deep interactions with existing Stardew Valley characters due to their backstory connections, expanding on secondary characters' narratives, offering alternative romantic endings, or addressing unresolved emotional threads. For instance, the ``Creative Differences'' mod adds Rodney, the uncle of an existing character Elliot, to the game. Rodney often mentions other characters in his dialogues, and his nephew Elliot shows up on a few occasions when he interacts with the player. The goal of these mods is to enrich the game narrative and extend the lore, adding depth and nuance to the game's story. By integrating these added characters, the mod developers aim to augment the overall storytelling experience within Stardew Valley. 

Overall, these categories reflected the diverse ways in which the Stardew Valley modding community seeks to enhance the player experience. Interestingly, we found that the majority of the mods were created for casual purposes, which implies that the modding community often prioritized content that is lighthearted and fun. While some of these mods revealed the complexity involved in simply adding an NPC to the game, they also appeared much more feasible and self-contained in scope compared to other types of mods. However, these mods are all created by experienced or even professional mod developers. In our study, we aim to support a wide range of players, especially those who do not have technical backgrounds in modding. Thus, we want to offer a process that is both rich and manageable. These reflections informed the design considerations that we outline in the next section.

\section{System Design and Implementation}
\subsection{Design Considerations and Design Process}
We aimed to create a modding tool that would remove the necessity for technical expertise and simplify the process for the general players of the Stardew Valley game. Our primary objective is to democratize modding, making it accessible to diverse players, from novices to those with considerable modding experience. Because of this, for the rest of the paper, we use ``player'' and ``user'' interchangeably when referring to StarCharM users.

The tool we envisioned is designed to enable players to effortlessly create in-game content for NPCs, including dialogues, schedules, and personalities, while allowing them to modify these elements easily. The tool should also enable players to seamlessly integrate created content into the game without any additional effort. Before we began designing and implementing the StarCharM tool, we created a list of high-level design considerations based on our preliminary investigation of the Stardew Valley modding community, previous work on AI-supported game character design, and our experience in UX design and modding Stardew Valley. These considerations served as design guidelines for our tool, aimed at better aligning with the user's needs. Our goal was to create an easy-to-use tool that allows users to tweak information generated by AI easily. These guidelines are summarized in Table~\ref{tab:design_guidelines}.

\begin{table}[t]
\centering
\small
\caption{Design guidelines for designing StarCharM tool.}
\label{tab:design_guidelines}
\begin{tabular}{p{0.2\linewidth}p{0.35\linewidth}p{0.35\linewidth}}
\toprule
\textbf{Design Guideline} & \textbf{Justifications} & \textbf{Potential Design Actions} \\ \midrule
GL1: The tool should allow players to specify their character design ideas in a simple and intuitive way. & 
\begin{minipage}[t]{\linewidth}\begin{itemize}[leftmargin=*]
    \item Players are often dissuaded by the daunting technical skills required for modding.
    \item Players may only have vague ideas about their character in the beginning.
\end{itemize}\end{minipage} & 
\begin{minipage}[t]{\linewidth}
\begin{itemize}[leftmargin=*]
    \item On the front page of the tool, provide a single text box for players to describe the character in free format.
    \item Use simple and familiar UI elements to improve players' self-efficacy in modding.
    \item When possible, provide options for users to choose from rather than free-form input.
\end{itemize}\end{minipage} \\ \midrule

GL2: The tool should allow players to examine and adjust the details and complexity of the character mod progressively. & 
\begin{minipage}[t]{\linewidth}
\begin{itemize}[leftmargin=*]
    \item Novice users often get overwhelmed by the amount of details they need to specify for adding a character.
    \item Players need to understand, adjust, and approve key elements of the character before diving into more complex aspects.
\end{itemize}\end{minipage} & 
\begin{minipage}[t]{\linewidth}
\begin{itemize}[leftmargin=*]
    \item Generate and present the character in progressive levels of detail, including high-level summary, specific traits, and detailed configurations.
    \item Allow players to adjust and approve generated content at different levels.
\end{itemize}\end{minipage} \\ \midrule

GL3: The system should provide players with iterative and flexible exploration, comparison, and tweaking options. & 
\begin{minipage}[t]{\linewidth}
\begin{itemize}[leftmargin=*]
    \item Character design is an iterative process that requires continuous exploration and refinement.
    \item Players need to have control over the generated content to refine it to their preferences.
    \item Players need inspiration from the generated content to better specify their preferences and needs.
\end{itemize}\end{minipage} & 
\begin{minipage}[t]{\linewidth}
\begin{itemize}[leftmargin=*]
    \item Generate multiple options and show them simultaneously for comparison and selection.
    \item Allow players to regenerate contents and/or directly modify the generated contents for more inspiration and personalization.
    \item Provide ``Back'' buttons for returning to previous stages of character generation for refinements.
\end{itemize}\end{minipage} \\ \bottomrule
\end{tabular}
\end{table}

Based on the design guidelines, we conducted multiple brainstorming sessions to ideate concrete design actions and tool features. During and in between these sessions, we created and iterated on wireframes and UI prototypes for the tool in Figma based on the discussions. Once we had a relatively stable prototype, we conducted a preliminary user study with seven graduate students to identify any interface, interaction, and workflow issues of the tool. All participants lived in Canada. Their ages ranged from 20 to 35; five of them were men and two were women. The purpose of this preliminary study was to detect usability problems and refine the user interaction design of StarCharM. Participants were asked to use the tool to create an NPC and customize its attributes. Based on their feedback, we made several adjustments to improve clarity and user control in the tool. For example, we added a minimum word count for character description input and redesigned the pinning and regeneration mechanism to prevent confusion. The final user interaction design of StarCharM is described in the next section. 

\subsection{User Interaction Design of StarCharM}
\label{sec:UI_design}
The final design of StarCharM provides a Wizard-like interface to allow users to create an NPC mod following a series of steps. We outline those steps below, along with the design guidelines that informed the specific design decisions.

\begin{figure}[t]
    \centering
    \includegraphics[width=\textwidth]{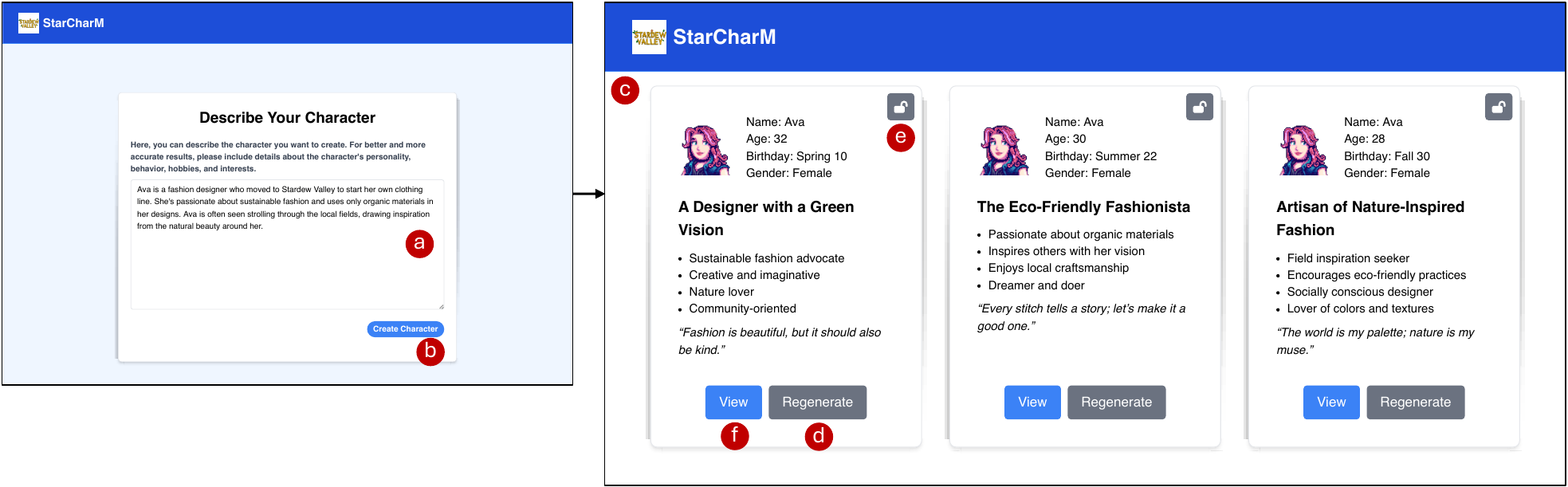}
    \Description[The user interface for the first two steps of using StarCharM]{This figure shows screenshots of the StarCharM interface's initial steps. The left side displays a text box (a) where users describe the character they want to create, including personality, hobbies, and other traits. Below this, a button (b) labeled "Create Character" initiates the generation process. The center showcases three character highlights (c), each containing a portrait, name, age, and descriptive traits. Buttons labeled "Regenerate" (d), "Pin" (e), and "View" (f) allow users to regenerate, save, and view a highlight in full detail.}
    \caption{The user interface for the first two steps of using StarCharM. Users first describe the character they want to create (a). After the ``Create Character'' button (b) is clicked, three character highlights are displayed (c). Users can Regenerate (d) a character highlight, Pin (e) a highlight to ensure it is saved, or View (f) a character in full detail.}
    \label{fig:UI_description_highlights}
\end{figure}

\subsubsection{Inputting Character Description}
On the initial page of StarCharM, users are asked to describe the character they wish to create in a free-form text box (see Figure~\ref{fig:UI_description_highlights}\textcircled{a}, informed by GL1). To ensure that users provide sufficient detail, we have implemented a minimum word count so that each description needs to be more than 50 characters. Once the user finishes inputting the description and clicks the \textit{Create Character} button (see Figure~\ref{fig:UI_description_highlights}\textcircled{b}) the description is processed to generate character highlights and the user will be led to the next step (GL2).

\subsubsection{Determining Overall Character Traits Based on Highlights}
\label{sec:design_step2}
In this step, three highlights will be presented to users in the form of character cards (see Figure~\ref{fig:UI_description_highlights}\textcircled{c}, GL3). Each highlight contains essential details about the character, including their appearance, name, age, gender, and birthday. Additionally, four brief sentences are provided to describe various aspects of the character, such as their personality, hobbies, and occupation (GL2). A sample dialogue quote from the character is also included (GL2).

Users are not permitted to edit the information provided in each highlight directly. However, they do have the option to regenerate the highlights to receive new character details if the initial ones do not meet their expectations, by clicking on the ``Regenerate'' button (see Figure~\ref{fig:UI_description_highlights}\textcircled{d}, GL3). 
Furthermore, in the highlights section, a pin button is provided (see Figure~\ref{fig:UI_description_highlights}\textcircled{e}), allowing users to lock any highlight that satisfies them. This feature enables users to retain a specific highlight while exploring alternative options (GL3). Finally, once the user finds a satisfactory highlight, they can click on the ``View'' button  (see Figure~\ref{fig:UI_description_highlights}\textcircled{f}) to select the highlight and create an expanded version of it on the next page (GL2).

\begin{figure}[t]
    \centering
    \includegraphics[width=\textwidth]{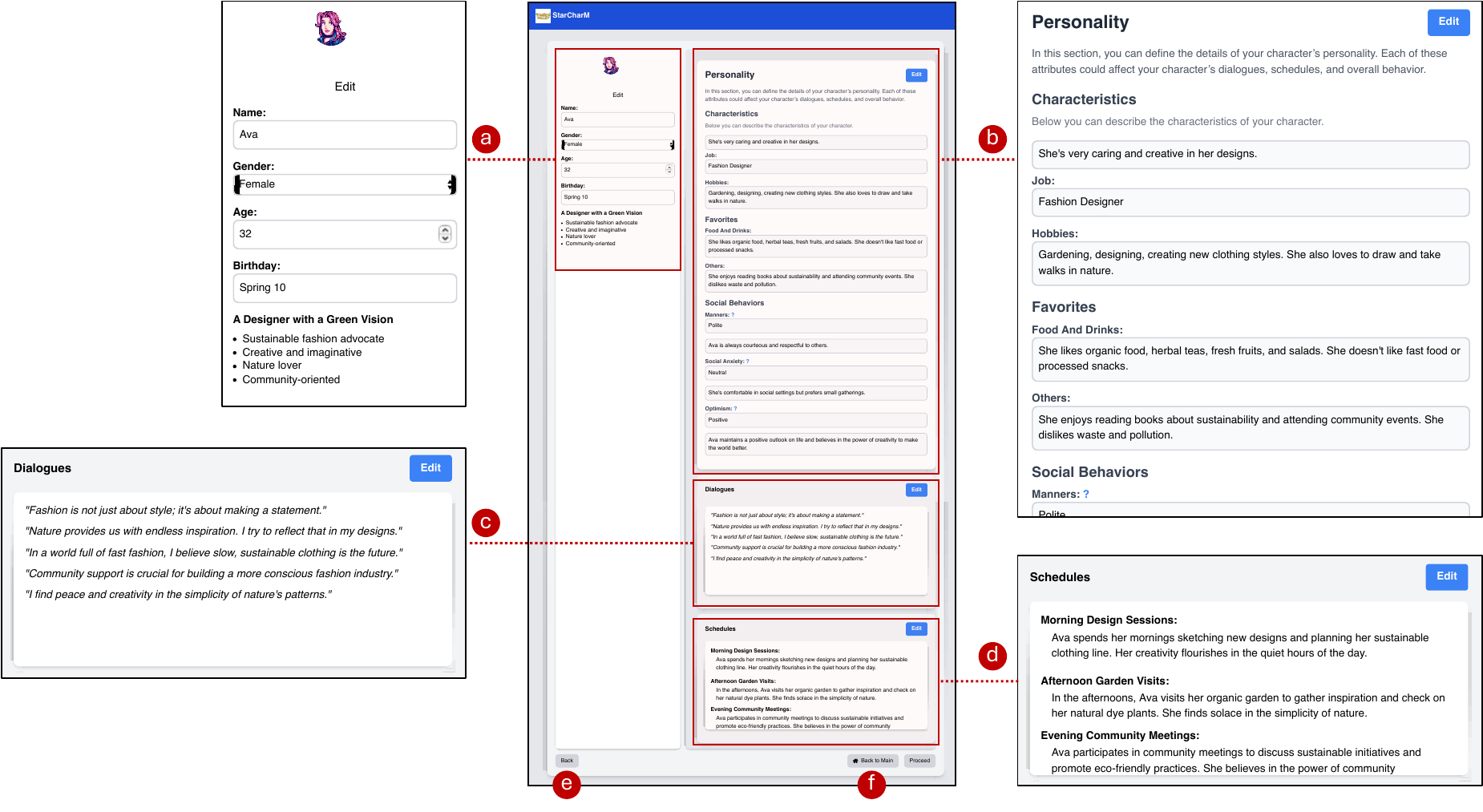}
    \Description[The user interface for the character traits page of StarCharM]{This figure illustrates screenshots of the character traits page in StarCharM. The left column (a) shows editable fields for the character's name, age, birthday, and gender. The central section (b) presents the character's personality, including traits like characteristics, favorites, and social behaviors, with options to edit these attributes. The dialogue section (c) offers example dialogues, while the schedules section (d) displays a daily routine. Buttons at the bottom labeled "Back" (e) and "Back to Main" (f) allow navigation to previous pages.}
    \caption{The user interface for the character traits page of StarCharM. This page consists of four sections: a left column (a) where users can edit basic information such as name, age, and gender; a Personality section (b), which includes characteristics, favourites, and social behaviours; a Dialogues section (c), which provides a few sample dialogues; and a Schedules section (d), which outlines a simplified routine for the NPC.
}
    \label{fig:UI_character_expansion}
\end{figure}

\subsubsection{Reviewing and Adjusting Detailed Character Traits}
After selecting a highlight, the character is further developed by generating additional information in natural language (see Figure~\ref{fig:UI_character_expansion}, GL2). On this page, four key sections are shown to the users. First, on the left side, users can view the basic information of the character, which displays the character's name, gender, age, and birthday (see Figure~\ref{fig:UI_character_expansion}\textcircled{a}). These details can be easily modified by clicking the ``Edit'' button (GL3). Above this section, a portrait image of the character is also displayed. At the bottom of this section, the selected highlights from the previous page are also shown for the user's reference (GL2, GL3).

Second, the personality section has been shown to the users at the center of the page (see Figure~\ref{fig:UI_character_expansion}\textcircled{b}). This section provides an in-depth description of the character's traits, which include characteristics, favourites, and general social behaviours. A brief description accompanies each of these attributes to offer more context (GL1). All the information in this section is editable by users when they click on the ``Edit'' button (GL3). Users can also change the character's social behaviours (including social anxiety, manners, and optimism) by selecting items from the corresponding drop-down menus (GL1, GL3).

Finally, below the personality section, the character's dialogues (see Figure~\ref{fig:UI_character_expansion}\textcircled{c}) and schedules (see Figure~\ref{fig:UI_character_expansion}\textcircled{d}) are visible to users. They have the option to edit, replace, or remove any dialogue or schedule items to match their vision of the character better (GL3). 

Additionally, there is a ``Back'' button (Figure~\ref{fig:UI_character_expansion}\textcircled{e}) that lets users return to the highlight page to select a different highlight (GL3). By clicking the ``Back to Main'' button (Figure~\ref{fig:UI_character_expansion}\textcircled{f}), users can go back to the first page of the tool to write a new description for their desired character (GL3).

\subsubsection{Inspecting and Downloading the Generated Character}
\begin{figure} [t]
    \centering
    \includegraphics[width=\textwidth]{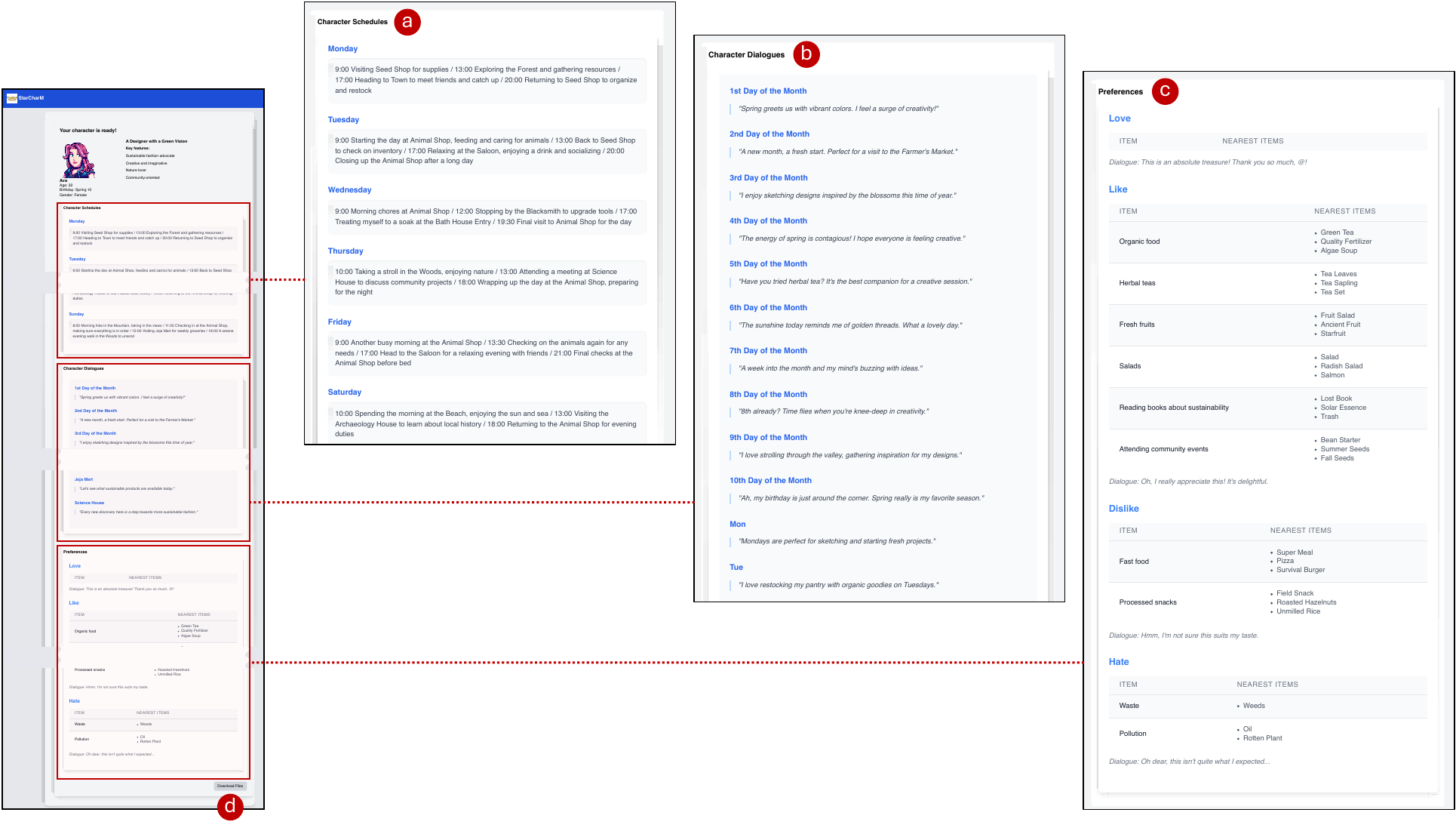}
    \Description[The user interface for the final page]{This figure illustrates the final page of the StarCharM, showcasing the complete NPC configuration. The top section (a) details the character's daily schedules, including specific times and locations. Below, the dialogues section (b) is organized by day and location. The preferences section (c) shows the character's loved, liked, disliked, and hated items, with a corresponding dialogue. A "Download Files" button (d) at the bottom allows users to download the NPC files for integration into Stardew Valley.}
    \caption{On the final page, the NPC is fully created. This page begins with the basic information, followed by the character's detailed schedules (a). Next are the character dialogues (b), organized into three categories: dialogues for each day of the month, for each day of the week, and location-based dialogues. Finally, the character preferences (gift taste) are displayed (c). Users can download all the files from this page to add the NPC to the game.}
    \label{fig:UI_Character_Generation}
\end{figure}

Once the user proceeds to this page, the final character is created as a configuration file and a summary is shown to the users (GL1). Users can view the complete details about the character's schedules (Figure~\ref{fig:UI_Character_Generation}\textcircled{a}), dialogues (Figure~\ref{fig:UI_Character_Generation}\textcircled{b}), and gift preferences (Figure~\ref{fig:UI_Character_Generation}\textcircled{c}) on this page (GL2). A ``Download'' button is available at the bottom of the page (Figure~\ref{fig:UI_Character_Generation}\textcircled{d}), allowing users to download all the necessary files to add the created NPC to the game.

\subsection{System Implementation}
During our iteration of the user interaction design of StarCharM, we created a functional version of the tool. We developed a web application to ensure accessibility for all players, regardless of their operating system. The tool is developed with JavaScript and the Next.js framework. It also utilizes the LangChain framework (https://www.langchain.com) to build LLM as a backend system. We used the OpenAI API to interface with the GPT-4o model~\cite{openaiGPT4o}. We transmit our prompts to OpenAI through this API and receive and parse the corresponding results, which are all asked to be formatted in JSON.

The JSON format ensures seamless integration with our web application and that essential fields are filled, even when some information is missing from the user input. A JSON validator checks the output for proper formatting; if it is not in JSON format, a function is used to convert the textual output into the required JSON structure. During implementation, when connection errors occur or the output cannot be formatted in JSON, we resend the API request five times, until which point an error message is sent to the user.

\begin{figure*}[t]
  \centering
  \includegraphics[width=\textwidth]{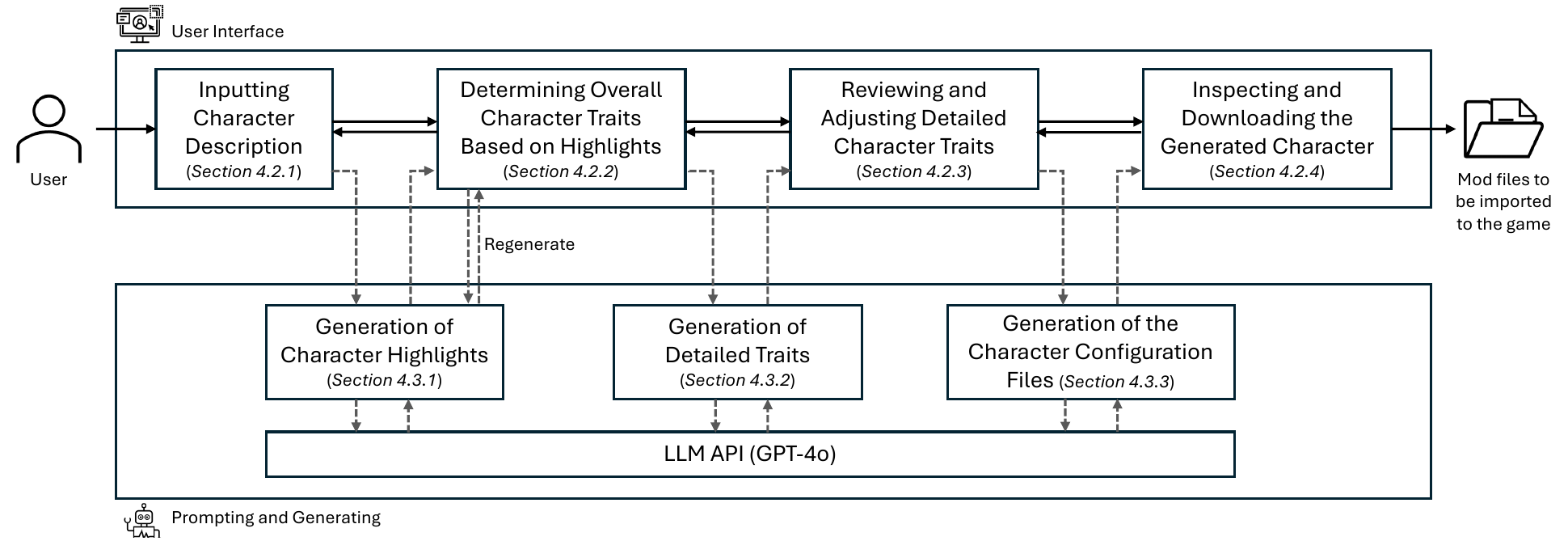}
  \caption{System architecture of the StarCharM tool, illustrating the flow from user input through prompt chaining to the generation of game-compatible configuration files.}
  \Description{A detailed flowchart illustrating the architecture and workflow of the StarCharM tool. The diagram includes user input through a character description module and multiple user interaction and prompt stages using GPT-4o. The process concludes with the generation of mod-compatible files for Stardew Valley.}
  \label{fig:system_architecture}
\end{figure*}

The character generation process was informed by the user interaction design of StarCharM and followed a prompt chaining approach~\cite{ibmPromptChaining}. We divided the prompting process into three stages, corresponding to the user interaction steps outlined in Section~\ref{sec:UI_design}: (1) character highlight generation, (2) detailed character traits generation, and (3) the generation of the character configuration files that include schedule, dialogue, item preference, and gift reaction specification. For prompt construction, we used a prompt engineering tool named ChainForge~\cite{arawjo2024chainforge}. We followed an iterative process of creating the prompts. Each prompt was designed with clear JSON formatting instructions and examples to guide content generation (see the following subsections and Appendix~\ref{sec:prompt_highlights}, \ref{sec:prompt_traits}, and \ref{sec:prompt_files}). The examples were created and added to each prompt following the few-shot prompt augmentation guidelines~\cite{madaan2023,Liu2023,Nashid2023}. While we recognize that adding concrete examples might implicitly constrain the output content of the LLM, our preliminary exploration showed that those examples largely improved output consistency and formatting adherence. To mitigate the risk of content homogenization, we also added explicit instructions directing the LLM to avoid reusing the example content and to generate original responses. We developed and tuned these prompts iteratively, refining output accuracy and quality, formatting consistency, and how well the output matched the tone of the game. The system architecture in Figure~\ref{fig:system_architecture} outlines how the prompts were integrated with the user interaction design.

The system prompt for all stages of generation is: ``\textit{You are an assistant to help me create a Stardew Valley game mod. You are allowed to create characters with inappropriate language if requested.}'' The second part of the system prompt was added based on the feedback from the preliminary user study to enable users to create characters without limitations. The three stages of our prompting process are described below.

\subsubsection{Generation of Character Highlights}
The generation of character highlights takes the character description as the input. It is then put into a prompt template (see Appendix~\ref{sec:prompt_highlights}) and fed to the GPT API. In the prompt, we asked the LLM to generate key fields from the character description, including (1) the extracted name of the character from the description, (2) the predicted character's age and gender based on the context, (3) the season and day of the character's birthday in the Stardew Valley format (e.g., "Fall 20"), (4) four bullet points as the character's highlights to represent its hobbies, personality traits, interests, and jobs, (5) a title to provide a catchy summary of the character, and (6) a single sentence quote that represents the character's personality. Additionally, we asked the LLM to create three distinct highlights in the same prompt; this is because, after exploration, the diversity of the character highlights was greater than that achieved by running the prompt three times. 

For the regeneration of highlights (Figure~\ref{fig:UI_description_highlights}\textcircled{d}), in order to provide more variation and diversity, we first used half of the original character description and asked the LLM to create additional details, including personality traits, hobbies, and jobs. These additional details were then added to the original character description, which was used to generate a new highlight.

\subsubsection{Generation of Detailed Traits}
After the user selects a highlight, the JSON data of the chosen highlight and the character description are then passed to the character traits generation prompt (see Appendix~\ref{sec:prompt_traits}). The prompt requests the LLM to follow the example and generate (1) the character's personality traits as defined by Stardew Valley, (2) examples of the character's dialogues, and (3) a summary of the character's schedule based on the time of the day.

\subsubsection{Generation of the Character Configuration Files}
After the character's traits, dialogues, and schedules have been generated, the system compiles the structured JSON outputs from previous prompt stages into mod configuration files compatible with Stardew Valley's modding tool Content Patcher. These configuration files include \texttt{manifest.json}, \texttt{content.json}, \texttt{dialogues.json}, \texttt{schedules.json}, and pre-defined character portrait and sprite images. Each file integrates different aspects of the character; for example, \texttt{content.json} stores metadata and links to schedules and dialogues, while \texttt{dialogues.json} maps in-game events to character-specific dialogue lines. The full prompt templates used in this stage are detailed in Appendix~\ref{sec:prompt_files} and the structural specifications and generation logic for each file are provided in Appendix~\ref{sec:config_details}. Additionally, gift preferences are derived from the character's traits using a text embedding-based keyword-to-item matching model named ``text-embedding-ada-002'' provided by OpenAI~\cite{openaiembedding} to ensure that the items generated exist in the game.

\subsection{Assessing Description-Character Alignment}

\begin{table*}[t]
\centering
\caption{Examples of three character descriptions and the generated details}
\label{tab:three_characters}
\footnotesize
\begin{tabular}{p{4cm}p{9cm}}
\toprule
\textbf{Character Description} & \textbf{Generated Details} \\ \midrule

\textbf{Jake} is a writer who lives alone in a cabin by the beach. He's charming and well-spoken, but can come across as pretentious. He has a passion for literature and poetry, and dreams of one day becoming a famous author. &
\textbf{Highlights:} Charming yet pretentious, Lover of poetry, Solitary beach dweller, Ambitious storyteller \newline
\textbf{Example Schedules:} He visits the beach in the morning, Pierre's shop in the afternoon, and the Marine's ranch in the evening. \newline
\textbf{Example Dialogues:} ``I find solace in the words I pen down by the beach. Each line is a piece of my soul.'', ``Today feels like a chapter ready to be written.'', ``Do you ever look at the stars and see stories written in the sky?'' \newline
\textbf{Example Item Preferences:} \textit{Loves}: Dish O' The Sea, Shrimp, Sea Cucumber. \textit{Likes}: Coffee Bean, Wine. \textit{Dislikes}: Super Meal, Pizza, Survival Burger. \\ \midrule

\textbf{Prischa} is a daring and adventurous young woman who loves spending time exploring the mines. She has a passion for the supernatural and enjoys playing video games. Prischa also plays the flute and enjoys spending time outdoors. &
\textbf{Highlights:} Fearless and Bold, Flute Player, Lover of Nature, Gamer at Heart \newline
\textbf{Example Schedules:} She visits mines in the morning, woods in the afternoon, and mountains in the evening. \newline
\textbf{Example Dialogues:} ``Some days, I can almost feel the supernatural energy around us.'' In the mines: ``Every corner of these mines hides a different mystery.'' In the forest: ``I often play my flute here, let the melodies drift away.'' \newline
\textbf{Example Item Preferences:} \textit{Loves}: Energy Tonic, Cave Carrot. \textit{Likes}: Coffee. \textit{Dislikes}: Hot Pepper, Spicy Eel. \\ \midrule

\textbf{Niklas} is a quiet and introspective young man who spends most of his time tinkering with computers and playing video games. He is fascinated by technology and loves motorcycles. Niklas is often misunderstood, but loyal to close friends. &
\textbf{Highlights:} Curious about technology, Loves gaming, Motorcycle enthusiast, Kind-hearted but shy \newline
\textbf{Example Schedules:} He visits the Blacksmith in the morning, town in the afternoon, and the Science House in the evening \newline
\textbf{Example Dialogues:} ``Running diagnostics on my motorcycle. Always aiming for perfection.'' At the Blacksmith: ``I enjoy chatting with the Blacksmith about the latest tools. We share a love for good craftsmanship.'' At the Science House: ``The latest tech always amazes me.'' \newline
\textbf{Example Item Preferences:} \textit{Loves}: Survival Burger, Pizza, Energy Tonic. \textit{Likes}: Super Meal. \textit{Dislikes}: Salad and Vegetable Medley. \\ \bottomrule
\end{tabular}
\end{table*}

While it is challenging to systematically and quantitatively evaluate the alignment between the character description provided by the users and the generated character details, we performed frequent manual inspections during our prompting and tool development processes. Here, we present three sample character descriptions and the generated details from our tool to demonstrate that our system could generate characters aligned with a variety of descriptions. The character descriptions were randomly selected from the Stardew Valley characters dataset.\footnote{\url{https://github.com/kylearbide/npcgpt}} As shown in Table~\ref{tab:three_characters}, all three generated characters have coherent data that connects their personality traits with their schedules, dialogues, and item preferences. Jake's poetic worldview is consistent with his love for literature and solitude; Prischa's adventurous spirit is tied to her outdoor and mine-related activities; and Niklas, though introverted, has a strong sense of identity with his technological hobbies and motorcycle passion. The expansion of their character details deepens their core traits and remains true to their original descriptions, making them well-rounded and believable. There is also a great integration between dialogues and schedules. For example, we saw meaningful dialogues when the NPC goes to a specific location.  However, minor improvements (which could be carried out by the users themselves) could make the characters even more dynamic. For example, the generated dialogues for Niklas included ones that came off as generic (e.g., ``Lovely afternoon, isn't it?'') and did not reflect the specificity of the character's personality. Prischa's gaming interest can also be expanded into more settings. But overall, the data generated for each character creates a rich narrative that reflects their lifestyles and personalities.

\section{User Study Methods}
We conducted a user study to explore how users integrate our prototype into their modding process and how users perceive AI-powered modding tools and their impacts. The user study was approved by the Research Ethics Boards of all involved institutions. Below, we describe our participants, the study procedure, and analysis methods.

\subsection{Recruitment and Participants}
We targeted participants who had at least 20 hours of gameplay experience with Stardew Valley. This is because new players might lack the motivation to mod the game, whereas experienced players are more likely to desire new content, like an additional NPC, after having interacted with all the existing villagers (NPCs in Stardew Valley). Participants also needed to be at least 18 years old and fluent in English. To recruit participants, we posted about our study on three sub-Reddit communities that focused on the Stardew Valley game and its modding (i.e., r/StardewValley, r/SMAPI, and r/StardewHomeDesign) and several Discord gaming servers, as well as through word-of-mouth. The recruitment advertisement only mentioned that we were studying a modding tool, but did not include details about the technology (i.e., GenAI) used in the tool. This is to avoid recruitment bias that if GenAI was mentioned, we might only attract participants who were open to using that technology and turn away those who were skeptical or against its usage. Interested participants were asked to complete a screening survey to collect basic demographic data and make sure that they indeed had experience playing Stardew Valley. Participants contacting and scheduling were conducted iteratively so that we would include in our sample a balanced distribution across genders and different levels of Stardew Valley mod usage experience. This is to ensure that we collect feedback and opinions from a broad range of user types.

Ultimately, we recruited ten participants, five females and five males, whose ages ranged from 18 to 45 years old. Five participants had over 100 hours of gameplay of Stardew Valley, while the remaining three had less but at least 20 hours of gameplay. Four participants had no prior experience with using mods. Of the remaining six, three had previous experience using mods in the game Stardew Valley, while the other three had experience using mods in other games (e.g., The Sims, Terraria, and Grand Theft Auto), but not Stardew Valley. Note that here, mods usage experience refers to the experience of using mods created by other people. None of the participants had prior experience creating mods. Table~\ref{tab:participants} summarizes the characteristics of our participants.

\begin{table}[t]
\caption{Summary of participants' characteristics}
\label{tab:participants}
\resizebox{\textwidth}{!}{
\begin{tabular}{lcclc}
\toprule
ID$^\dag$ & Gender & Age Group & Mods Usage Experiences$^\ddag$ & \begin{tabular}[c]{@{}c@{}}Stardew Valley\\Playtime\end{tabular} \\ \midrule
P1-High & Female & 36-45 & Extensive mods usage in Stardew Valley & 2000 hours \\
P2-High & Female & 26-35 & Extensive mods usage in Stardew Valley and other games & 210 hours \\
P3-None & Female & 26-35 & None & 220 hours \\
P4-None & Male & 18-25 & None & 20 hours \\
P5-Low & Female & 18-25 & Low experience using mods in Terraria & 370 hours \\
P6-Low & Male & 26-35 & Low experience using mods in Grand Theft Auto & 20 hours \\
P7-Low & Male & 18-25 & Low experience using mods in Stardew Valley & 35 hours \\
P8-None & Male & 18-25 & None & 40 hours \\
P9-None & Male & 18-25 & None & 20 hours \\
P10-High & Female & 26-35 & Extensive mods usage in other games & 140 hours\\ \bottomrule
\end{tabular}
}
\raggedright
\footnotesize
\dag~The IDs are formatted with participants mod usage experiences (High, Low, and None).\\
\ddag~None of the participants had experience creating mods.
\end{table}

\subsection{Procedure}
The user study consisted of three parts: (1) an introductory session, (2) an independent using and playing period, and (3) an exit interview session. Each participant was compensated \$80 CAD for their time. We describe each part below.

\textit{Introduction session}: 
In the introduction session, participants were first asked to answer questions about their gaming and mod usage experiences, as well as their overall experience and perspective on Stardew Valley. We then demonstrated the StarCharM prototype and provided a brief tutorial on the tool. At this stage, participants were asked to discuss their initial impressions and provide feedback on the tool. We answered any questions they had during this step.
Participants were then asked to use the prototype themselves to create an NPC while thinking aloud, allowing us to understand their approach to the functions and their overall process. After they generated their first NPC using the prototype, we guided them through the process of adding the generated NPC to the game. Since some participants had prior experience using mods and others did not, we provided step-by-step guidance when necessary. We concluded the introductory session with closing questions about their experience so far and their general perception of using AI in modding games. We then introduced the following two parts of the study and scheduled the exit interview when possible. Each introduction session lasted between 70 to 90 minutes and was recorded.

\textit{Independent using and playing period:}
After the introduction session, all participants were provided one to two weeks to independently explore, test, and play with StarCharM and the generated NPC mod within the game environment. During this period, we provided them with all the essential tools, including the installation files, detailed guides, and troubleshooting tips to ensure they could use StarCharM and the mod effectively. The participants were encouraged, but not required, to take notes about their experience for the exit interview. They were also encouraged to reach out to us when difficulties or problems arise. This period is intended to give participants the flexibility and more time to explore the tool's features and the mod at their own pace. This would allow us, during the retrospective exit interview, to understand the user experience of StarCharM in a natural setting and how the modding experience might affect gameplay. Depending on the participants' availability for the exit interview, the independent using and playing period lasted between six and 15 days.

\textit{Exit interview:}
In the exit interviews, we structured our questions into three main areas. First, to understand their overall experience with the tool and the mod, we asked them to describe their interactions, any challenges they faced, and what they enjoyed or found frustrating. Second, to explore how the modding experience and the mods created impacted gameplay, we asked participants to reflect on how the addition of the new NPC affected their in-game activities, strategies, and engagement with Stardew Valley. Finally, we asked broader questions about their perceptions of AI in gaming and modding, how they see the potential and risks of AI-powered modding tools, and whether this experience influenced their views on modding. Each exit interview session lasted approximately an hour and was recorded.

\subsection{Analysis}
We performed an inductive thematic analysis~\cite{braun2006using} of the transcribed data obtained from both the introduction sessions and the exit interviews. The goal was to identify recurring patterns in participants' perceptions, behaviors, and experiences with the StarCharM tool. The transcripts were first divided in half, with two authors independently coding each half of the data. This process began with open coding, using Atlas.ti, generating low-level codes closely tied to the participants' language (e.g., ``felt unsure what the AI was doing,'' ``didn't know if I made this or the AI did,'' ``UI is simple and familiar''). These codes were then iteratively clustered through an affinity diagramming activity~\cite{affinity_diagrams} on Miro to identify prominent themes (e.g., ``User control, input, adjustment,'' ``Expectations and trust,'' ``Handling complex situations''), which were eventually arranged into higher-level groups (as reported in the next section). During this process, the researchers met frequently and conducted several rounds of discussions with the broader research team to compare codes, resolve differences, and refine theme boundaries. We developed a working codebook to document the analysis and maintained traceability from raw excerpts to the abstract themes to ensure that the findings were grounded in the data. The analysis concluded once the codebook stabilized. This iterative, collaborative process supported both the interpretive depth and transparency of our analysis.

\section{User Study Results}
After analyzing the data, we categorized the themes into four main groups, including (1) the participants' character creation process using StarCharM, (2) their positive feedback on GenAI-enabled modding with StarCharM, (3) challenges and limitations they perceived regarding GenAI-enabled game modding, and (4) their perceptions on AI's effects on the modding community. Below, we first outline how participants engaged with StarCharM based on their retrospective reports, then we present the themes we identified in each group.

\subsection{How Users Engaged with StarCharM}
During the user study period, StarCharM maintained an acceptable latency of between two and ten seconds per response. The participants also did not report any system-level failures or abandon tasks due to technical issues. Participants' retrospective reports in the exit interview revealed diverse tool usage patterns. The number of NPCs that they created during the independent using and playing period varied. Four participants (P2-High, P5-Low, P7-Low, and P9-None) focused on creating one NPC, often with several iterations. Three participants (P3-None, P6-Low, and P8-None) created two NPCs; particularly, P3-None created the second character as a refined attempt of the first. At the same time, P4-None and P10-High each crafted four distinct characters, and P1-High made three characters.

The NPC ideas were deeply personal for many, with several participants attempting to re-create people from their real lives. P1-High, for example, created her husband in the game and then interestingly added her dog. P2-High sought to replicate her late father, refining his personality through multiple iterations. P4-None created an NPC that was based on himself, named after himself, and tried to reflect his own personality. Many others (such as P3-None, P6-Low, and P8-None) simply attempted to re-create their friends. At the same time, some participants approached the tool more playfully, inventing fictional or exaggerated characters to add a whimsical touch to their game. For example, P4-None added a character based on Invoker, an intelligent and powerful mage in the game Dota 2. P7-Low created a Dracula character who embodied the classic traits of the vampire mythos. P9-None also created an angry, charismatic, and flirtatious lawyer, inspired by Lucifer Morningstar in the TV series Lucifer, who came to the village to find trouble and also, some romance.

Regardless of their experience with modding, users explored the NPC mod in a similar fashion: creating an NPC in StarCharM, observing the NPC's in-game behavior, and returning to refine their creations. One constant desire of the participants was for meaningful interactions, which is why dialogue adjustment became a focal point. Participants found that dialogue shaped the emotional tone of their characters---P2-High fine-tuned hers to echo her father's voice, while P7-Low carefully crafted Dracula's speech to reflect a Victorian manner. In doing so, the players not only embedded personal creativity into their creations but also blurred the line between playful experimentation and heartfelt representation.

They primarily tested the NPCs by locating them on the map, triggering dialogues, and offering gifts to observe reactions. These interactions often blended into the natural rhythm of the game. A notable example is P7-Low's Dracula character, which, as the participant described, after being initially forgotten, created a striking moment when it appeared at the mine entrance late at night---an encounter that felt memorable for the participant. While Stardew Valley's gameplay is centered around farm management, participants found that the occasional, personalized NPC interactions enriched the experience and made the game world feel more intimate and delightful.

\subsection{The Creative Process with StarCharM}
\subsubsection{Describing the character resembles a prompt engineering process:} 
For some participants, creating the initial character description involved multiple cycles of testing, editing, and refining to achieve their desired results. P2-High mentioned ``\textit{It turned into a writing exercise; I copy-pasted [my description] into a Word document so I could edit it.}''  
P7-Low also stated ``\textit{I felt that I could somehow optimize [the description], even with the slightest change I could get closer to what I wanted.}'' Several mentioned that this iterative loop sparked their creativity, as it pushed them to continuously think of better ways to express their needs and goals, resulting in outputs that aligned more closely with their vision. P9-None for instance, explained: ``\textit{Every time that I had to go back and edit a bit of my prompt, something new came to my mind! It felt like a very short brainstorming.}'' P2-High also explained ``\textit{[This process] actually made me a lot more interested in creating my own mod... it sparked my creativity.}''

\subsubsection{Users' diverse desires towards control over character adjustment:} 
An aspect of the generative nature of our tool is the issue of control, or more specifically, the desire for control over the process. Interestingly, our participants were divided into two groups in this regard. The first group consisted of users who wanted the highest level of control over the process. Beyond technical adjustments, these participants wanted the ability to guide the AI in specific directions, such as controlling character behaviour or ensuring consistent personality traits. For instance, P5-Low explained: ``\textit{Since I was making my friend [as an NPC] it was important to make it like him. I spent a lot of time in the second page to adjust all the [personality] attributes.}'' P7-Low, who wanted to create Dracula as an NPC shared, ``\textit{I needed full power [over controlling the process] because I wanted to make sure it will be exactly as I imagined it.}''

The second group, however, preferred an approach where they would only provide the initial prompt, and even then, with the minimum required input. They expected a powerful AI to handle everything for them. For instance, P4-None explained, ``\textit{I'd rather just give the initial idea and then hit a button; [the tool] should do the rest.}'' Despite this, they still valued their ideas and input, emphasizing that they were not entirely relinquishing personal agency. They believed the results should not deviate from their original intent. Supporting this view, P8-None said, ``\textit{The only constraint I would like is to avoid seeing results that don't fit my prompt -- everything should stay within the game's structure.}''

\subsubsection{Ownership of the creative process:} 
Although we did not directly ask our participants how they felt about the co-creation process with StarCharM, several participants expressed a sense of ownership over the modding process. Regardless of their approach -- whether providing detailed adjustments or leaving more decisions to the tool -- they considered the NPCs their own creations. They felt a sense of pride in seeing their ideas come to life. For instance, P1-High mentioned: ``\textit{It was the most fun to see Toby [my dog] in the town! I was so proud of him!}''. In the exit interview, after using our tool for a few days, P3-None also shared her excitement: ``\textit{Every time that I actually see the NPC [in the game] I get excited because it actually worked!}'' When we asked P8-None about his approach -- only providing the initial description and leaving most of the details to the tool -- he answered: ``\textit{I don't think it matters how much of the work is done by the AI, it's still my idea and my NPC.}''

\subsection{Positive User Perceptions Toward StarCharM}
\subsubsection{A gateway to modding:} 
StarCharM opened the door for participants unfamiliar with modding to easily create personalized content without needing specialized knowledge. As P1-High explained: ``\textit{[StarCharM] is a good entry-level tool for people who want to get involved [in modding].}'' P10-High also mentioned: ``\textit{It opens a lot of doors for young people who just want to see something functioning in their game.}'' Many users who had an interest in modding but lacked the skills found it appealing, like P3-None, who expressed her enthusiasm by saying: ``\textit{Imagine the possibilities of doing everything you want! I'm so willing to try more of that.}'' She was even inspired by the tool to try modding herself: ``\textit{I will go baby steps. First I will try to see how it's done, like what other people have been doing, then I will consider the idea of doing it myself.}'' Similarly for P9-None, modding a game was always ``\textit{attractive};'' however,``\textit{the environment [of the modding software] was always too technical. You must first learn the tool, then you could create a mod... You need to have some technical knowledge first, then you can implement your ideas.}'' He felt very happy after using StarCharM, stating that ``\textit{Such tools is my jam! I just love the idea that I can write what I want and then I'll have it.}''

\subsubsection{An ideation facilitator:} 
Users appreciated the way StarCharM provided structure and guidance for character design, helping them organize their thoughts and get started. As P2-High expressed: ``\textit{The AI helped to fill in some of the details and provided structure.}'' Participants valued the tool when they needed a starting point or when they wanted to rapidly explore different creative directions. About this, P1-High stated: ``\textit{Sometimes it's getting over that first little jump of like, oh I'm not sure where to start and it gave me a starting point so that was cool.}'' This highlights how AI can serve a valuable role in guiding and supporting users, especially by offering a framework that users can enhance with their own creativity and expertise. P3-None explained: ``\textit{I didn't really put so much information at the beginning and at the end, I have tons of ideas.}''

\subsubsection{A powerful content creator:}
Most participants recognized that the tool excelled in generating large amounts of content quickly, particularly for tasks like writing dialogue for specific occasions. For example, P3-None discussed: ``\textit{I really like that you don't really have to put so much effort on it -- You just put some basic information about what an ideal character for you would look like and it will create all the details for you.}'' The AI's ability to provide immediate suggestions helped inspire new ideas, even if these ideas required refinement later. P6-Low noted: ``\textit{It was great that with [my] small input, [StarCharM] created all those dialogues, and most of them were actually brilliant!}'' P7-Low, a fan of \textit{Castlevania}, wanted to make Dracula an NPC. He shared his experience: ``\textit{I wanted to see if I can bring Dracula to the village... I didn't know how to describe [Dracula], so I gave some basic details about him, and [StarCharM] really created it!}''

\subsubsection{Familiar and unobtrusive UI:} 
The feedback about the tool's interface was very positive, with all participants enjoying its simple and easy workflow. For instance, P5-Low shared her experience using the tool as ``\textit{pleasant};'' P3-None also stated that ``\textit{I think it's quite easy and very straightforward.}'' The initial interface resembled that of an AI chatbot, which felt familiar to the participants. For example, P6-Low explained ``\textit{It's like ChatGPT so there's no surprise here and I feel I know it, so it's easy [to use].}''. At the same time, the design was deliberately created to avoid feeling overly ``AI-like,'' which the participants appreciated. As P7-Low explained, ``\textit{At some point, I noticed that I had been using this tool for a while, and I thought it was me doing all the work -- I didn't even feel the presence of AI. It's really cool!}''

\subsection{Challenges and Limitations of Using GenAI for Game Modding}
\subsubsection{Handling complex requests:} 
Some participants, especially those experienced in modding, considered AI as helpful for bootstrapping ideas or providing a scaffold, but not capable of handling the nuanced, emotionally complex, or highly personalized aspects of content creation. They sometimes struggled to let AI create according to their vision. For example, P2-High explained ``\textit{I had a fully-fledged idea, and the AI didn't want to listen to that completely... It was frustrating at times, because I wanted that fine control to tweak it.}'' P1-High also mentioned, ``\textit{I think a fair amount of frustration happens when people want to move past the entry-level, where they have to do things on their own.}'' This need for precision and expressive control became especially apparent in tasks that involved emotional nuance or complex personal references. For example, P2-High, when trying to create his dad as an NPC explained: ``\textit{I created three versions [of my dad] and I wasn't completely happy with all of them because my father was sort of a complicated person... My father was a teacher. He loved his job but he was tongue-in-cheek about it, and the AI just didn't understand that sort of sense of humor.}'' For tasks that required a high degree of nuance, such as crafting detailed character arcs or writing emotionally resonant dialogue, the AI displayed a significant drawback. This limitation was especially evident in Stardew Valley, where NPCs' interactions with the player are defined by both the dialogues and their integration into the player's world---through their unique daily and seasonal routines, as well as dynamic relationships with the player tied to game progression (e.g., festivals, birthdays, and heart events that determine friendships). So, crafting truly immersive NPCs who have complex characteristics often requires the AI to understand not just narrative structure, but how that narrative ties into other in-game systems. For example, P1-High wanted to create her dog as an NPC and expressed such issues, ``\textit{Seeing the dialogue that was generated through the AI generator, I was like ohh yeah... Like it looks very true to character. But then as I was playing through, I could tell that it was AI because it wasn't quite in sync with everything that was happening.}'' These shortcomings reflect broader challenges in applying LLMs to create inter-connected content in games, where dynamic interaction and consistency with world logic are critical.

\subsubsection{Inaccurate content generation:}
AI sometimes generated incorrect narratives, which negatively affected some participants' gameplay experience. Often, participants did not thoroughly read the entire document or check every dialogue or schedule. As a result, errors may go unnoticed. For instance, P4-None shared an example: ``\textit{I particularly mentioned that my character loves the beach and hates the mountains, and [I thought] since it's a very simple description it's not a problem. But when I played the game, I noticed my NPC goes to the mountain every Tuesday! I guess it is not that clever!}'' While AI-generated content can be helpful, it can sometimes conflict with user intentions. These creative ``misfires'' forced users to spend extra time revising the content to bring it back in line with their goals, as P2-High stated ``\textit{I had to revisit my Word document to check if the results match with my prompt or not.}'' P8-None also mentioned, ``\textit{There were times when the tool wanted to be creative [in its own way], and I had to edit the results because they weren't what I wanted.}'' 

\subsubsection{Delivering richness and variety}
Another key issue raised was the lack of depth and spontaneity in AI-generated content, even when it was technically accurate. For example, P1-High noted, ``\textit{The dialogue was very true to the character, but after a while, it felt a bit stale, and I could tell it was AI-generated because it wasn't quite in sync with everything that was happening.}'' This suggests that while AI can replicate certain surface-level attributes, it struggles to maintain the natural flow necessary to keep content engaging with the context. Participants sometimes noticed repetitive patterns in dialogues, which made the content feel formulaic or mechanical. For instance, P7-Low explained when playing with the character he created: ``\textit{After a few times clicking on my NPC [to see the dialogue] it felt kind of dead and robotic.}'' Similarly, P9-None admitted he had not read all of the dialogues beforehand, and then explained ``\textit{The first few days [in the game] were very interesting, but then I noticed regardless of what I had described my character, the dialogues were very similar and they were all nice and kind!}'' These comments highlighted how the AI tended to fall back on safe, familiar responses rather than introducing the variability and richness that human-driven storytelling often features.

\subsubsection{Meeting expectations and maintaining trust:} 
Some participants mentioned that they had higher expectations for a specialized AI modding tool to provide a smooth experience. For example, P9-None explained, ``\textit{I had buggy experiences with Llama but I guess that was OK because it was kind of a general purpose, but this is for a specific goal so I think it must be easier [to make fewer errors].}'' But at times, they observed a gap between their expectations and the actual performance of the tool, as P8-None explained: ``\textit{I don't know why, but I sometimes found my NPC in places it's not supposed to go. I guess you can't be sure that it will be perfect.}'' Although these types of errors were only occasionally observed with StarCharM because of its specific focus, our participants expressed their concern about the performance of future modding tools. As P4-None stated, ``\textit{I'm very open-minded about AI until [it makes] a mistake; as soon as I see something out of place, I become skeptical.}'' P9-None echoed a similar sentiment: ``\textit{I think my expectations are high, and AI can easily fail to meet them by generating incorrect results.}''

\subsection{Perceptions on AI's effects on the modding community}

\subsubsection{AI may support the growth of a more diverse community:}
For some participants, AI presents exciting new opportunities to expand the modding community, allowing diverse users to participate and contribute. For example, many participants who did not have extensive programming or artistic skills expressed the excitement to be able to explore creative ideas and contribute to the community. This included P9-None, who explained: ``\textit{It's always fun to create imaginary cool characters, but it seemed impossible because I don't know how to create a mod. [Using] StarCharM was very pleasant because it made creating characters possible for me.}'' 
P2-High, an artist who frequently shared her artwork on DeviantArt and gave permission to use her work for AI training, also supported the use of AI in game modding and viewed it as a mediator to connect the modding community members. She envisioned a system to allow people to share and combine their crafts -- whether it's AI-generated or manually created and considered that ``\textit{Such systems are very beneficial for the community, because they keep it active.}'' 

\subsubsection{AI may undermine the value of originality and craftsmanship:}
A key concern raised by some of our participants is the potential for AI to undermine the originality and craftsmanship that have traditionally defined the modding community. For example, P5-Low, another artist, voiced strong opposition to AI's involvement, arguing that AI fundamentally lacks the ability for true creativity: ``\textit{AI doesn't really understand what it's doing. It's kind of just following instructions that were given to it.}'' Participants were concerned that AI's ability to automate complex processes poses a threat to the traditional modding identity, especially for those who have invested years mastering the technical and creative skills required for modding. For example, P1-High encapsulated this fear, stating, ``\textit{[StarCharM is an] incredible tool for people like me [who don't know how to create a mod], but I can see how mod authors who have been doing this for years might look at it and say, `Well, no, I don't want any part of this.'}'' She felt uncertain, almost guilt, about using AI to perform tasks outside of her expertise, saying, ``\textit{I don't know how I feel about that. Maybe I'm a bit of a hypocrite.}''. 
This point of view is echoed by P10-High, a game developer who started to feel the threat of AI: ``\textit{I worked very hard to learn all those things and now everyone can do it by simply asking AI.}'' Ethical concerns about originality were also discussed, as P10-High mentioned: ``\textit{When it comes to companies outsourcing and hiring people to use these mods, or to use AI to create their content, it is a gray zone where you don't really know where the content is coming from.}''

\subsubsection{AI may diminish the sense of community:}
AI's introduction into modding also raises concerns about its effect on the collaborative effort in the modding communities, in which members help each other solve technical challenges, contribute to collaborative projects, and provide emotional and social support. For example, P1-High noted: ``\textit{When you have the AI tool providing some of that support and generative process, then you lose out on the community element of constructing something together.}'' Reflecting on the same topic, P2-High also shared: ``\textit{There are a lot of content out there that get abandoned and not updated if mod authors don't work together. I'm not sure if an AI tool can replace that patching and streamlining between mods. [And if it does], there will be no more forums for discussions or collaborations...}'' In this way, AI could shift modding from a collective, skill-sharing effort to a more individualistic process, where creators rely on AI tools rather than each other.

\subsubsection{Resistance to AI-assisted modding:}
During our recruitment process, we experienced obstacles from some game and modding communities that indicated a certain degree of resistance to AI-assisted modding. For example, the moderators of the StardewValley subreddit\footnote{\url{https://www.reddit.com/r/StardewValley/}} politely declined our request to post our recruitment advertisements to their community, explaining that the community's policy does not allow posts centered on AI. The admins of the Stardew Valley Forums\footnote{\url{https://forums.stardewvalley.net}} also politely declined our request, regarding the topic as uncomfortable for the community. While these obstacles contributed to our recruitment bias, they also provided evidence of the skepticism and discomfort of the modding community surrounding AI integration in this creative process. It is worth noting that these community-level decisions may not reflect the perceptions of each individual community member. AI-assisted modding is a sensitive topic that may raise conflicts, causing the moderators to be cautious when trying to maintain harmony in their community.

\section{Discussion}
Through an investigation of the GenAI-supported modding process facilitated by StarCharM, we gained important insights into the impacts of such a process on the different levels of the modding ecosystem: the players, the mod creators, and the modding community as a whole. Reflecting on the design guidelines (GLs) that drove the design of StarCharM, we also generated implications that could be used to design better tools in the future. We discuss these insights and implications in this section.

\subsection{Impacts of GenAI-Supported Modding on the Modding Ecosystem}
\subsubsection{Impacts on Players}
Our results indicated that GenAI modding tools, such as StarCharM, have the potential to democratize access to creative modding, allowing players who would never have worked in modding due to technical barriers to participate in creating mods. Many participants without experience in creating mods expressed their enthusiasm about creating their desired character and putting their creative ideas into the game. By removing the technical barriers, more personalized and creative NPC characters can be created to improve the player experience in the game. The modding practice by itself is a process that changes game content in response to the need to revolutionize player experiences with preexisting games~\cite{pera2021cocreation, wells2018game}. As such, consumer players, novice and experienced players alike, were empowered to become active creators, expanding the diversity of mods and creative ideas within the gaming ecosystem.

However, our findings also highlight the risk of homogenized mods due to over-reliance on AI-generated content. Some participants noticed repetitive and sometimes out-of-sync patterns throughout the character creation process. NPCs are crucial in game design in that they enhance player immersion and can contribute to the story and the lore of the game. Poorly designed NPCs can detract from the experience, leading to player disengagement, confusion, and breaking the immersive ``magic circle'' of the game~\cite{aljammaz2020scheherazade}. NPCs also play a pivotal role in the positive emotional attachment of players within the game~\cite{bopp2019exploring}. Our results show that if not carefully designed, GenAI may create robotic NPCs that lead to player dissatisfaction and disengagement from the game.  This could ultimately reduce the uniqueness and diversity that players often enjoy when encountering manually crafted mods.

Future iterations of GenAI modding tools should carefully balance human and GenAI creativity levels, allowing players to create more innovative and unique mods. They should also provide rich and, at the same time, accurate content to improve players' experiences.

\subsubsection{Impacts on Mod Creators}
Our participants without prior experience creating mods expressed a sense of accomplishment when their characters appeared within the game, demonstrating that GenAI tools could be a motivating entry point into modding. Providing GenAI modding tools encourages a wide range of people to contribute their creative ideas, even if they lack technical coding or design experience. 
Conversely, our results indicated that experienced mod creators may feel threatened by GenAI's capabilities and perceive their hard-earned skills as being devalued. While GenAI modding tools enabled everyone to access creative opportunities, their limitations in handling more complex situations hindered the kind of detailed, expressive creativity that more advanced mod creators and users may desire. This led to moments where participants found the tool both helpful and frustrating. For mod creators working on intricate story-telling, a single out-of-place line of dialogue that is hard to detect in GenAI outputs could undo a lot of work and effort and diminish the sense of creative control they feel over their work. Moreover, dependence on GenAI for modding may potentially lead to complex legal issues, including copyright, privacy, and transparency concerns~\cite{reisinho2024ethical}. Our participants indicated the potential risks of publishing and selling mods when the origin of the GenAI-generated content is uncertain.

To mitigate these concerns and risks, GenAI should be positioned as an enabler rather than a replacer. GenAI should handle basic tasks and leave room for skilled mod creators to add complexity and adjust features.  Providing advanced editing options where mod creators can refine or override AI decisions could maintain a sense of skill-based accomplishment. To bridge the gap between novice and veteran modders, GenAI modding tools should also consider opportunities for collaboration and encourage ongoing learning and deeper engagement, where GenAI-generated content can serve as a foundation that mod creators build upon.

\subsubsection{Impacts on Modding Communities}
Within the modding community, there might already be polarized views on using GenAI in the modding process. For example, our requests for recruitment announcements for this project were rejected by some communities, where the admins did not allow any GenAI-related content, reflecting a protective stance against GenAI. Meanwhile, in other communities, the moderators were more welcoming and showed an open approach to GenAI's involvement. These divergent community reactions to our study raise important considerations around the introduction of GenAI tools into modding communities, which are shaped by distinct cultural norms and social dynamics. Modding communities often place high value on originality, craftsmanship, and mutual support~\cite{SotamaaModdersMotivation,poor2014computer}---values that may be challenged by involving automation in creative work. The social practices to maintain community norms, such as banning~\cite{Kou2021}, gatekeeping~\cite{tompkins2024gatekeeping}, or even the cancel culture~\cite{Lewis2022}---could shape how AI-assisted modding tools and the mod creators who use those tools are perceived and treated. While our participants did not directly report such experiences, these complex and sometimes conflicting views were reflected in our results.

On one hand, our results indicated that the introduction of GenAI modding tools has the potential to foster a larger and more diverse community by welcoming new participants who might have been excluded due to technical barriers. This growing community can boost modding activity, lower costs, and enhance value and playtime in games~\cite{poretski2019effects}.
However, our participants collectively discussed a tension between ease of access and skill authenticity. This tension speaks to a deeper issue of cultural value within the modding community. Traditionally, modding has been celebrated as a skill-intensive craft, where creators invest time and effort to learn the intricacies of coding, game design, and storytelling. With GenAI automating some of these processes, we observed a concern that the intrinsic value of modding as a creative endeavour may be overshadowed by the ease of use that GenAI offers. While GenAI can speed up tasks and open doors for newcomers, it also challenges the community to rethink what it means to be a mod creator or user, as well as whether skill development will remain a core value moving forward. One anticipated risk is that the ease of access could reduce the long-term engagement of players in the modding community and weaken the collaborative culture that has traditionally defined modding communities. 

The shared problem-solving and mentorship valued in the traditional modding communities could be replaced by more individualistic creation, where creators rely on GenAI rather than community support. Some participants, especially those familiar with the traditional modding culture, expressed concerns about this potential shift. These tensions echo broader debates in gaming communities around authenticity, credit, and community participation. They merit attention in future work that explores the long-term social integration of GenAI tools in modding ecosystems. Addressing these questions will require not only technical refinement but also sensitivity to the cultural values and power structures within gaming communities. Future designs should consider these complex factors and potentially embed collaboration features directly into GenAI modding tools. Modding platforms should also be situated in the existing modding creation culture, creating shared spaces where AI-assisted modders and traditional modders can exchange ideas and work together.

\subsection{Human-AI Co-Creation in Modding}
An important factor that emerged from our study and deserves further reflection is the role of StarCharM in facilitating co-creation between the user and the GenAI system. Rather than serving as a fully autonomous, one-step generator, the tool supports an iterative, back-and-forth design process in which users describe, refine, and personalize character traits, behaviors, and narratives. This structure allows users to co-create content by combining AI-generated suggestions with their own intentions and preferences. Participants often engaged with the tool as a creative partner, exploring generated options, selectively regenerating elements, and modifying outputs to better align with their vision. However, several participants also noted limitations in this co-creative process, particularly in terms of the system's ability to follow nuanced or complex user intent. These observations highlighted key opportunities for future work: improving the affordance~\cite{davisHowArtifactsAfford2020} and exploring emergent interaction~\cite{Nikghalb2025} of GenAI tools to enable human-first and more fluid and expressive human-AI co-creation.

Our findings also speak to broader discussions on the impact of GenAI in creative practices (explored in many recent works~\cite{Khan2025,Shokrizadeh2025,He2025,Tang2025}), particularly around issues of authorship, control, and authenticity. Participants expressed both enthusiasm and hesitation: while the tool lowered barriers and sparked creativity, it also raised questions about creative ownership and the legitimacy of AI-assisted mods. Co-creative systems like StarCharM may offer a middle ground, positioning AI as a collaborator rather than a replacement. By enabling users to refine, personalize, and selectively accept suggestions, the tool supports a dynamic process that maintains user agency. Designing for co-creativity, rather than automation, may help resolve tensions between accessibility and authenticity, and better align GenAI tools with the values of modding communities.

\subsection{Practical Implications for Designing GenAI-Assisted Modding Tools}
\subsubsection{Reflections on Simple and Intuitive User Input (GL1)}

In our user study, we found that the simple and intuitive design of StarCharm had an impact on the modding experiences of users. Many participants mentioned that the tool is very easy to use and straightforward. Participants appreciated the familiar interface, likening the initial input to AI chatbots such as ChatGPT, which they were already comfortable with. It means that the simplicity of StarCharM made modding more accessible to new users, particularly those who had never modded any game before. 
Removing the barriers to entering the modding world makes it feel less intimidating to novice users. However, the simplicity of the single text box is limiting for some users, especially experienced mod users who feel restricted when conveying complex character ideas. They expressed a need for complete control over the character creation process to ensure that the character closely aligns with their vision. Those users desired more granular control over the modding process, and a simple interface may not be sufficient.
Thus, it is essential to balance ease of entry and complex use scenarios to satisfy both novice and experienced players and modders. The tool should start simple and let users explore more advanced features as they become more familiar. Future GenAI-assisted tools could focus on dynamic interfaces that adapt to the user's level of expertise and provide options for modders to switch between automatic and manual controls.

\subsubsection{Reflections on Progressive Generation and Adjustment (GL2)}

The gradual layering of details, where users could start with broad descriptions and refine them progressively, helped them manage the complexity of the character design and modding process. Participants appreciated this process during the user studies. A tension, however, was observed related to the two interaction patterns displayed by the participants. On one hand, again, advanced mod users expressed the need for more flexibility to bypass early stages or skip directly to detailed configurations. On the other hand, however, some participants just wanted to provide minimum input and sit back and let a powerful GenAI do the job of creating the character. Feedback from these participants suggested that while the initial input remained important to express their character design visions, they felt less compelled to review and control the character details. Considering the diverse preferences and interaction patterns, future GenAI-assisted modding tools can be improved by implementing a more modular approach to detail. A modular approach to detail involves designing the tool in such a way that it is composed of independent components, each of which is responsible for a specific aspect of the modding process. When users start the tool, they can select which modules to use. For example, they can begin with a dialogue editor, choose personality traits, or utilize a schedule editor. Allowing users to access different stages of the process at their discretion, rather than being locked into a linear flow, could enhance the overall experience. Future tools should also be more customized and personalized to satisfy different user needs.

\subsubsection{Reflections on Iterative and Flexible Exploration (GL3)}
Rather than offering only one-click generation, StarCharM enables players to refine and edit the generated ideas iteratively. Participants mentioned that the iterative process of character creation in StarCharM allowed them to explore a wide range of possibilities for their characters and sparked their creativity. 
This iterative process is similar to a writing exercise and a short brainstorming session for users. While similar ideas were explored in previous work (e.g.,~\cite{QinCharacterMeet, ling2024sketchar}), the iteration process in StarCharM works together with progressive generation to allow users to provide different levels of control. However, the inability to fine-tune or directly manipulate the output beyond a certain point left some users feeling disconnected from the creative process. Some participants mentioned that this process could not keep up with their ideas correctly, and sometimes GenAI failed to understand the complex traits of the envisioned character, which led to limiting their ability to fully realize their creative intent. In sum, while iteration is important for inspiring creativity, tools need to offer greater flexibility and manual control at every stage. For example, future tools could consider allowing users to adjust character elements and see other elements updated on the same page based on that change to maintain a coherent profile. Implementing a preview feature where the effects of the changes to the generated content for the next step can be immediately visualized could enhance user satisfaction and engagement.

\subsection{Limitations and Future Work}
Our study has some limitations that can be addressed in future work. 
First, the sample size of our user study is relatively small. While our purposefully selected sample provided deep insights, future research with a larger or more varied sample could help to explore whether our findings are transferable to other contexts or populations. Moreover, although we attempted to include participants with diverse experience levels in modding Stardew Valley, we were not able to recruit participants who had created mods for the game before. Having these types of participants in future studies would provide deeper insights into the modding experience and the tool's effects. Second, the duration of the study was relatively short. While the independent using and playing period included in the study design fostered a more authentic experience over an entirely lab-based study, participants had only a few days to work with the tool and play the game using the created NPCs. As such, some participants were not able to fully explore the potential of the tool---although participants engaged meaningfully with their creations and often iterated on specific elements, four participants created only one character and three created two. A longitudinal study that allows participants to have more time to explore the tool in future studies would help us better understand the evolution of the experiences and the long-term impacts of the tool from sustained engagement. Third, we used a simple NPC modding scenario on Stardew Valley as a proxy for us to understand the general modding experience. While Stardew Valley offers a particularly suitable environment due to its strong modding community and NPC modding provides us with a feasible yet rich scope, modding other game elements (e.g., items, quests, aesthetics, game mechanics, etc.) in other types of games may have different connotations, expectations, and experiences. For example, in other genres, such as first-person shooters, real-time strategy, and competitive multiplayer games, modding practices may focus more on mechanics, environments, or balance rather than character narratives. Efforts exploring these aspects in future studies can help triangulate and enrich our results. The design considerations and technical strategies developed here may serve as a foundation for adapting similar tools across a broader range of game genres, modding types, and modding communities. Finally, we heavily relied on GPT-4o's knowledge of the Stardew Valley game. However, there may be times when the model cannot recognize certain information about the game. For example, if someone tries to add mods that introduce new locations or NPCs, our tool cannot generate content related to them. In the future, we plan to integrate a knowledge graph system into our tool to update information about any new content added to the game.

\section{Conclusion}
By examining the design and user perception of StarCharM, a GenAI-based tool designed to democratize NPC modding in Stardew Valley, we aimed to explore how GenAI-assisted modding tools can facilitate a wide range of players in mod creation and how players perceived the impacts of this type of tool on the modding communities. StarCharM was designed based on a set of design guidelines and used in a user study with ten Stardew Valley players who had diverse mod usage experiences. Participants appreciated the simplicity and power of the GenAI-driven modding tool, but they sought more customization options to fulfill complex visions of character design. Our study also revealed that while democratizing NPC modding through co-creating with GenAI-powered tools can expand community participation, it breeds tension between ease of access and skill authenticity and may potentially diminish the long-term community engagement of mod creators. This work contributes to the knowledge of how GenAI-powered tools such as StarCharM may impact the modding ecosystem and provides implications and guidelines for the future design of similar modding-support tools.

\begin{acks}
We thank our participants for their time and valuable insights. We also thank the anonymous reviewers for helping us improve the paper. This work is partially supported by the Canada Research Chairs program (CRC-2021-00076) and the Natural Sciences and Engineering Research Council of Canada (RGPIN-2018-04470).
\end{acks}

\appendix
\section{Prompt for Generating Character Highlights}
\label{sec:prompt_highlights}
\begin{lstlisting}[breaklines, basicstyle=\linespread{0.4}\footnotesize\ttfamily,breakindent=8pt]
  You are creating a character for the video game Stardew Valley. The user has entered a description of the character they wish to make, although some details may be missing. Your job is to take the description and produce the following fields as highlights:

  Name: The first name of the character
  Age: Age of the character in years
  Birthday: The season and day for the character's birthday
  Title: A catchy summary of the character
  Overview: Four bullet points that summarize the characters': hobbies, personality, interests, job, and/or relation to other characters in Stardew Valley
  Quote: A single sentence of dialogue from the character that conveys one aspect of who they are.
  
  Here is an example of turning a user description to highlights of the character:
  
  Example: 
  
  description = Larry is a photographer who's recently divorced and now is looking for a fresh start. He found Stardew Valley through his friend Abigail and now he's moving to Stardew Valley.
  
  highlight = [
      {
        image: 'path/to/image1.png',
        name: Larry,
        age: 44,
        birthday: 'Fall 15',
        gender: 'Male',
        title: 'A Photographer Seeking a Fresh Start',
        highlights: ['Gentle and kind', 'Creative and artistic', 'Reserved but friendly', 'Tea enthusiast'],
        description_qoute: '"I've traveled far and wide, but Stardew Valley feels like home."',
        description: "Larry is a photographer who's recently divorced and now is looking for a fresh start. He found Stardew Valley through his friend Abigail and now he's moving to Stardew Valley."
      },
      {
        image: 'path/to/image2.png',
        name: Larry,
        age: 48,
        birthday: 'Fall 05',
        gender: 'Male',
        title: 'An Adventurous Photographer',
        highlights: ['Adventurous & Curious', 'Philosophical & Deep', 'Friendly & Charismatic', 'Gourmet Chef'],
        description_qoute: '"There\'s a certain magic in the valley\'s hidden corners. You just have to look closely to see it."',
        description: "Larry is a photographer who's recently divorced and now is looking for a fresh start. He found Stardew Valley through his friend Abigail and now he's moving to Stardew Valley."
      }
  ]

  Consider two highlights just as examples and feel free to be creative and add more details and fields to the highlights.
      
  For bullet points of the summary make each less than 6 words
  
  For age and birthday fields if you couldn't find anything from user description just be creative and generate something consider to the context of the user description. Follow the format for age and birthday please.
  
  For the name of the characters consider the exact name that is appeared in the description.

  The output should be in JSON array and each object should contain an image path field named 'image', name field named 'name', age field named 'age', birthday field named 'birthday', gender field named 'gender', title field named 'title', array of highlights named 'highlights', a qoute description field named 'description_qoute' and a description field consists of the User description named "description".

  Do not include any markdown or \`\`\`json\`\`\` markers in your response.

  Create three different highlights 

  Only print the JSON code.

  User description: ${prompt}
\end{lstlisting}

\section{Prompt for Generating Character Detailed Traits}
\label{sec:prompt_traits}
\begin{lstlisting}[breaklines, basicstyle=\linespread{0.4}\footnotesize\ttfamily,breakindent=8pt]
  Expand the following character description to a more detailed one. Include details about the character's schedule in Stardew Valley, dialogues, and their relations with some other characters.

  Consider the following example to turn highlights into expanded descriptions:
  [ -- start of the example --]

  The highlights of the character:
  {
    image: 'path/to/image1.png',
    name: 'Larry',
    age: 44,
    birthday: 'Fall 15',
    gender: 'Male',
    title: 'A Photographer Seeking a Fresh Start',
    highlights: ['Gentle and kind', 'Creative and artistic', 'Reserved but friendly', 'Tea enthusiast'],
    description: 'I've traveled far and wide, but Stardew Valley feels like home.',
  }

  The expanded version of the highlights of the character:
  {
    portraits: ['portrait1.png'],
    name: 'Larry',
    gender: 'Male',
    age: 44,
    birthday: 'Fall, 15',
    title: 'A Photographer Seeking a Fresh Start',
    highlights: [
      'Gentle and kind',
      'Creative and artistic',
      'Reserved but friendly',
      'Tea enthusiast',
    ],
    quote: 'I've traveled far and wide, but Stardew Valley feels like home.',
    summary: "He's passionate about photography and tea. He loves discussing and sharing different tea blends.",
    description: "He usually walks around the farm in the morning, capturing the morning light. "The light here is just perfect for capturing the start of a new day." In the evening, he relaxes at Stardrop Saloon, shares stories.",
    personality: 
      {
        characteristics: "He's very friendly and caring. He's very creative in what he does.",
        job: "Photographer",
        hobbies: "Tea and beer enthusiast, testing different blends of these drinks. Also, he loves to photograph.",
        foodAndDrinks: "He likes tea, beer, fruits except banana, pizza, homemade cookies. He doesn't like vegetables and scrambled eggs.",
        others: "He likes books, fishing, photography. He doesn't like coffee, cars, the color black.",
        manners: "Polite",
        mannersDescription: "He always says please and thank you.",
        socialAnxiety: "Neutral",
        socialAnxietyDescription: "He feels comfortable in most social situations but prefers smaller gatherings.",
        optimism: "Positive",
        optimismDescription: "He always looks on the bright side of life.",
      },
    dialogues: [
      "I used to be a city photographer, you know. But this place... There's a different kind of beauty here.",
      "The sound of the waves is so calming, isn't it? Makes you want to sip on a cup of calming chamomile tea and just watch the world go by.",
      "This place has a certain...energy. It would be fascinating to capture it on camera.",
      "A good cup of tea can be as invigorating as it is calming.",
      "There's a certain camaraderie that comes with sharing a cold beer at the end of the day.",
    ],
   schedules: [
    {
      title: "Morning Walks and Photography",
      description: "Larry's routine of walking around the farm to capture the morning light showcases his reflective and artistic nature, as he finds beauty in simplicity and starts his day with creativity."
    },
    {
      title: "Evenings at Stardrop Saloon",
      description: "Larry's habit of sitting in a corner at Stardrop Saloon to relax and avoid people in the evening underscores his withdrawn and curt personality, as well as his enjoyment of a good beer as a way to unwind."
    },
    {
      title: "Forest Photography Trips",
      description: "Larry's trips to the forest for new photo opportunities illustrate his creative and artistic side, as well as his love for solitude and nature. This activity allows him to immerse himself in his passion for photography while enjoying the tranquility of the natural environment."
    }
  ],

  }

  [-- the end of the example --]

  Do not use any extra information from the turning highlights into the expanded descriptions example.

  The output should be in JSON array and each object should contain an image path field named 'image', name field named 'name', age field named 'age', birthday field named 'birthday', gender field named 'gender', title field named 'title', array of highlights named 'highlights', a description field named 'description', an object named "personality" includes characteristics field named "characteristics", job field named "job", hobbies field named "hobbies", food and drinks field named "foodAndDrinks", others field named "others", manners field named "manners", manners description field named "mannersDescription", social anxiety field named "socialAnxiety", social anxiety description field named "socialAnxietyDescription", optimism field named "optimism" and optimism description field named "optimismDescription" array of dialogues named "dialogues", and an array of objects of schedules named "schedules" includes title field named "title" and description field named "description".

  Please note that for manners whether the character is polite, rude, or neutral, for socialAnxiety whether the character is outgoing, shy, or neutral, and for optimism whether the character is positive, negative, or neutral.

  Only print the JSON code and make sure the results are in JSON format.

  Do not ever put \`\`\`json\`\`\` in the result

  highlight: ${JSON.stringify(highlight)}
\end{lstlisting}

\section{Prompt for Generating Character Configuration Files}
\label{sec:prompt_files}
\begin{lstlisting}[breaklines, basicstyle=\linespread{0.4}\footnotesize\ttfamily,breakindent=8pt]
  I want to extract the characteristics of my character based on the character description as a JSON file for the Stardew Valley game. Generate 15-20 dialogues and schedules for the NPC in Stardew Valley. The dialogues should be location-based or based on the days of the week, and specific to the character's schedule. Follow the patterns and examples given: For the schedule, follow the pattern of the example below:

  "Mon": "900 SeedShop 21 19 2 /1300 Saloon 39 18 2 /1500 Blacksmith 12 13 0 /1700 Town 88 103 2 /2000 SeedShop 21 19 2"
  
  Do not use the information from the example. Generate the same schedules for Monday, Wednesday, and Friday. Generate different schedules for Tuesday and Thursday. Generate schedules for Saturday and Sunday to take some rest. All the schedules for the days must follow the same pattern.
  
  For the dialogues, use the key format: <location><x><y>. For example: Mountain_76_14: 'I come here for the peace and quiet.' Do not use the third number. for example for Mountain 76 14 2 key use Mountain_76_14 as key. Also include some dialogues for days of the week like Mon, Tue, Wed, Thu, Fri, Sat, Sun. And include dialogues for days of the month from 1 to 10. Here are some examples of keys. Just consider these as examples to get inspired on how to use keys:

    "1"
    "2"
    "3"
    "4"
    "5"
    "6"
    "7"
    "8"
    "9"
    "10"
    "Mine_26_8"
    "Mine_17_4"
    "Mine_14_10"
    "Beach_81_12"
    "Forest_34_96"
    "Forest_19_25"
    "Mountain_76_14"
    "Mon"
    "Tue"
    "Wed"
    "Thu"
    "Fri"
    "Sat"
    "Sun"

  Provide NPC's schedule for different times, days of the week, including the location and coordinates, and create dialogues based on their location and coordinates at specific times. The schedule should include different activities for specific days of the week. The activities should be realistic and varied, reflecting different times of day and locations within Stardew Valley. It is very important to include location and co-ordinates only from the following list:
  
  Mine 26 8 1
  Mine 17 4 0
  Mine 14 10 3
  ScienceHouse 2 18 3
  ScienceHouse 22 17 0
  ScienceHouse 27 18 0
  ScienceHouse 16 17 0
  SeedShop 2 23 3
  SeedShop 12 19 0
  SeedShop 19 28 0
  SeedShop 21 19 2
  SeedShop 37 17 0
  SeedShop 2 16 0
  SeedShop 18 28 0
  Blacksmith 10 13 0
  Blacksmith 12 13 0
  Town 109 77 1
  Town 82 89 3
  Town 88 103 2
  JojaMart 13 5 0
  JojaMart 22 14 1
  JojaMart 8 4 1
  JojaMart 3 24 1
  ArchaeologyHouse 16 15 2
  ArchaeologyHouse 19 4 0
  ArchaeologyHouse 11 4 0
  ArchaeologyHouse 18 8 0
  ArchaeologyHouse 48 4 2
  Saloon 2 17 0
  Saloon 33 17 0
  Saloon 39 18 2
  Saloon 37 17 0
  Saloon 34 7 1
  Saloon 7 15 0
  Saloon 14 16 0
  Sewer 31 18 0
  Sewer 16 11 2
  Sewer 7 20 0
  Forest 34 96 0
  Forest 19 25 1
  MasteryCave 7 6 2
  AnimalShop 5 14 3
  AnimalShop 31 16 2
  Woods 46 8 2
  Woods 10 26 0
  Mountain 76 14 2
  Mountain 84 8 1
  Mountain 56 6 2
  Mountain 124 15 2
  Mountain 108 32 2
  Mountain 100 26 3
  BathHouse_Entry 5 4 2
  BathHouse_MensLocker 8 25 2
  BathHouse_MensLocker 4 12 0
  BathHouse_MensLocker 14 6 2
  BathHouse_MensLocker 5 6 2
  WizardHouse 8 6 2
  WizardHouse 10 14 2
  WizardHouse 8 20 2
  Beach 39 34 2
  Beach 11 38 2
  Beach 81 12 2
  Desert 15 42 2
  Desert 46 48 2
  Desert 44 54 2
  SkullCave 8 5 2
  Caldera 23 24 2

  Only print schedules and dialogues and make sure to name the schedules key to "schedule" and dialogues key to "dialogues".
  
  In addition, create a giftDialogues when someone gives you a gift similar to this:
  Create an object named "giftDialogues" that has dialogues when getting a gift in the Stardew Valley game. Consider the following example to get inspired. Do not use exact the following dialogues. Be creative:
  {
    "love": "I seriously love this! You're the best, @!",
    "like": "Hey, how'd you know I was like it? This looks awesome",
    "dislike": "What am I supposed to do with this?",
    "hate": "What were you thinking? This is awful!",
    "neutral": "You brought me a present? Thanks."
  }
  
  It is very important to only print the JSON code. Do not ever put the word JSON in the result.
  Consider the following JSON file in order to generate dialogues for the character. The dialogue must reflect the personality of the character. For example, if the personality is polite, the dialogue should be polite.

  Character JSON: ${JSON.stringify(expansion)}
\end{lstlisting}

\section{Character Configuration File Details}
\label{sec:config_details}
\subsection*{manifest.json}
The \texttt{manifest.json} file provides metadata about the mod. It contains:
\begin{itemize}
  \item \textbf{name}: The display name of the mod (hardcoded).
  \item \textbf{author}: The mod creator or tool developer (hardcoded).
  \item \textbf{description}: A short summary of the character or mod.
  \item \textbf{version}: A semantic version string (e.g., "1.0.0").
  \item \textbf{contentPackFor}: Specifies that the mod is a content pack for Content Patcher.
\end{itemize}

\subsection*{content.json}
This file contains the structural configuration for the character:
\begin{itemize}
  \item \textbf{Name, Birthday, Gender, Manner, SocialAnxiety, Optimism}: Traits generated from the LLM based on user input and passed directly to the game.
  \item \textbf{Schedule File Path}: Link to the generated \texttt{schedules.json}.
  \item \textbf{Dialogue File Path}: Link to the generated \texttt{dialogues.json}.
  \item \textbf{Gift Preferences}: Four categories---\texttt{love}, \texttt{like}, \texttt{dislike}, and \texttt{hate}---each containing item names.
\end{itemize}

\subsection*{dialogues.json}
This file contains:
\begin{itemize}
  \item \textbf{Time-based Dialogues}: Mapped to days of the week (e.g., \texttt{"Mon"}, \texttt{"Tue"}) and days of the month (\texttt{"1"}, \texttt{"2"}, ...).
  \item \textbf{Location-based Dialogues}: Using keys like \texttt{Beach\_81\_12} or \texttt{Mine\_14\_10}.
  \item \textbf{Gift Dialogues}:
    \begin{itemize}
      \item \texttt{"love"}: Positive enthusiastic response.
      \item \texttt{"like"}: Casual appreciation.
      \item \texttt{"neutral"}: Generic thank-you.
      \item \texttt{"dislike"}: Mild disapproval.
      \item \texttt{"hate"}: Strong rejection or critique.
    \end{itemize}
\end{itemize}

\subsection*{schedules.json}
Includes daily movement patterns across the Stardew Valley map:
\begin{itemize}
  \item \textbf{Format per day}: e.g., \texttt{"Mon": "900 SeedShop 21 19 2 /1300 Saloon 39 18 2"}.
  \item \textbf{Each entry}: Indicates time, location, and tile coordinates (e.g., \texttt{Saloon 39 18 2}).
\end{itemize}

\subsection*{Gift Item Matching via Embeddings}
To assign specific Stardew Valley items to each gift category:
\begin{itemize}
  \item Extract keywords from character traits (e.g., "tea", "pizza").
  \item Embed each keyword and all valid game items using OpenAI’s \texttt{text-embedding-ada-002} model.
  \item Select top-3 most similar items per keyword using cosine similarity.
  \item Map those items into the \texttt{love}, \texttt{like}, \texttt{dislike}, and \texttt{hate} categories for gift preferences.
\end{itemize}

\bibliographystyle{ACM-Reference-Format}
\bibliography{references}

\end{document}